\documentclass[aip,jmp,superscriptaddress,longbibliography,
eprint,nofootinbib,
reprint % uncomment for two columns
]{revtex4-2}
\usepackage[utf8]{inputenc}
\usepackage[T1]{fontenc}
\usepackage{amsmath,amssymb,amsfonts,amsthm,mathtools,mleftright,bm}
\usepackage{dsfont}
\usepackage{graphicx}
\usepackage[dvipsnames]{xcolor}
\usepackage{hyperref}
\hypersetup{colorlinks=true, linkcolor=BrickRed, urlcolor=blue!50!black, citecolor=blue!50!black}
\usepackage{cleveref}
\usepackage{bigints}

\newcommand\FUB{\affiliation{Freie Universität Berlin, Department of Mathematics and Computer Science, Berlin, Germany}}

\newcommand\DIEP{\affiliation{Dutch Institute for Emergent Phenomena, 1090GL Amsterdam, The Netherlands}}

\newcommand\ZIB{\affiliation{Zuse Institut Berlin, Takustr.\ 7, 14195 Berlin, Germany}}

\newcommand\CORRAUTHOR{\email[Corresponding authors: ]{m.delrazo@fu-berlin.de, winkelmann@zib.de}}

\newcommand{\ord}[1]{^{(#1)}}
\newcommand{\dss}{\displaystyle}

\newcommand\contract[1]{#1}%\overline{#1}}
\renewcommand\phi\varphi

\begin{document}
	
\title{Chemical diffusion master equation: formulations of reaction--diffusion processes on the molecular level}

\author{Mauricio J. del Razo}
\CORRAUTHOR
%\email{m.delrazo@fu-berlin.de}
\FUB
\DIEP

\author{Stefanie Winkelmann}
\CORRAUTHOR
%\email{winkelmann@zib.de}
\ZIB

\author{Rupert Klein}
\FUB
% \email{rupert.klein@math.fu-berlin.de}

\author{Felix Höfling} 
\FUB
\ZIB

\date{\today}

\begin{abstract}
The chemical diffusion master equation (CDME) describes the probabilistic dynamics of reaction--diffusion systems at the molecular level [del Razo \emph{et al.}, Lett.\ Math.\ Phys.\ 112:49, 2022]; it can be considered the master equation for reaction--diffusion processes. The CDME consists of an infinite ordered family of Fokker--Planck equations, where each level of the ordered family corresponds to a certain number of particles and each particle represents a molecule. The equations at each level describe the spatial diffusion of the corresponding set of particles, and they are coupled to each other via reaction operators --linear operators representing chemical reactions. These operators change the number of particles in the system, and thus transport probability between different levels in the family. In this work, we present three approaches to formulate the CDME and show the relations between them. We further deduce the non-trivial combinatorial factors contained in the reaction operators, and we elucidate the relation to the original formulation of the CDME, which is based on creation and annihilation operators acting on many-particle probability density functions. Finally we discuss applications to multiscale simulations of biochemical systems among other future prospects.
\end{abstract}

\maketitle
\section{Introduction}

It is a well-established paradigm to consider biochemical dynamics as an interplay between the spatial transport (diffusion) of molecules and their chemical kinetics (reaction), both of which are inherently stochastic. There exist different approaches for modeling and mathematically formalizing such reaction--diffusion processes, ranging from reaction--diffusion master equations \cite{gardiner1976correlations,drawert2012urdme,isaacson2013convergent, winkelmann2016spatiotemporal,smith2019spatial}, where spatial transport is modeled by diffusive jumps between local compartments, to concentration-based approaches, such as deterministic \cite{grindrod1991patterns, murray2001mathematical,brauns2020phase,kostre2021coupling} or stochastic partial differential equations \cite{kim2017stochastic} and partial integro-differential equations \cite{isaacson2022mean}. 
The preceding modeling approaches may be regarded as approximations or limiting cases of particle-based reaction--diffusion (PBRD) models, which explicitly resolve the diffusive trajectories of individual particles in space and time, as well as reactions between them. In the standard PBRD models, particles move freely in space following Brownian motion, or any other form of diffusion process \cite{hoefling2013anomalous,weiss2014crowding}, and can undergo chemical reactions, which involve one, two or more reactants in such a way that the reaction rate can depend on the positions or relative positions between the reactants \cite{doi1976stochastic,smoluchowski1918versuch}. Because of their high complexity, PBRD systems are mostly studied numerically by means of Monte Carlo simulations of the underlying stochastic reaction--diffusion process.

The mathematical formalization and analysis of PBRD models, however, is difficult because reactions constantly change the number of particles of each species, changing the dimension and composition of the system. Recent work presents a probabilistic framework and the characteristic evolution equation for PBRD termed \textit{chemical diffusion master equation (CDME)} \cite{del2021probabilistic}. The CDME consists of an infinite ordered family of Fokker--Planck equations (i.e., an enumerated collection), where each equation corresponds to a certain number of particles $n=0,1,2,\dots$. The equations, for each fixed $n$, describe the spatial diffusion for the corresponding $n$-particle probability distribution, and they are coupled via reaction operators that express the changes in the system's state due to chemical reactions. These operators change the number of particles in the system, and thus they can be conveniently expressed in terms of creation and annihilation operators \cite{del2021probabilistic}, following a classical analogue of the quantum mechanical Fock space concept \cite{doi1976second,grassberger1980fock}. First steps towards solving the CDME analytically by means of the Malliavin calculus were taken recently \cite{lanconelli2022using}. A more comprehensive introduction on the topic can be found in ref.~\onlinecite{del2021probabilistic}.

In this work, we explore the CDME from several perspectives and present three approaches to motivate and formulate it. This work not only improves our understanding of how to formulate the CDME, but it also provides a more illustrative and accessible approach to practitioners than the original work \cite{del2021probabilistic}. In general, the CDME is composed of a diffusion operator and several reaction operators (one for each included reaction), all of them acting on a symmetric many-particle distribution function. 
In analogy to the well-known chemical master equation \cite{gillespie1992rigorous,mcquarrie1967stochastic,qian2010chemical,winkelmann2020stochastic}, which characterizes spatially well-mixed stochastic reaction kinetics, each reaction operator consists of a \textit{loss} term describing the probabilistic outflow from a given configuration state by the reaction, and a \textit{gain} term that captures the probabilistic inflow from other configuration states due to the reaction. The crucial part is to determine these loss and gain operators for different types of reactions in the absence of a spatially well-mixed setting; examples are binding and unbinding, creation and degradation, and mutual annihilation. Here, non-trivial combinatorial factors enter for preserving symmetry and normalization of the many-particle distribution functions under time evolution. The local rate function, which defines the probability per unit of time for a reaction to take place depending on the spatial positions of it's reactants and products, has to be transformed into an expression that takes the whole system state into account. This issue is addressed via the following three approaches:
 
 \begin{enumerate}
 	\item  We use the local rate functions to specify also the loss and gain operators on a local scale (acting on subsets of reactants and products), and then combine them into \textit{global} operators taking all combinations of reacting subgroups into account. The combinatorial factors included in the operators are motivated by an inductive argument. The CDME may then directly be expressed in terms of these \textit{global} loss and gain operators (\cref{sec:CDME}).

 	\item   The \textit{global} loss and gain operators are expressed in terms of \textit{many-particle propensity functions}, which define the probability per unit of time for a reaction to occur as a function of the whole system state. We explicitly derive these many-particle propensity functions from the given local rate functions using permutations and Dirac $\delta$-distributions. For the exemplary settings of decay and binding it will be shown that the resulting CDME agrees with the one of the first approach (\cref{sec:ManyParticlePropensity}).

 	\item  The operators in the CDME are expressed as expansions in terms of creation and annihilation operators as in ref.~\onlinecite{del2021probabilistic}. These expansions can be condensed in a compact notation that allows us to write the CDME, for a given system of reactions, in a simple, fast and straightforward manner. The combinatorial factors do not appear explicitly, instead they are naturally encoded in the creation and annihilation operators (\cref{sec:FockCDME}). A dictionary specifying the relation between the compact notation for the expansions and the concrete algebraic expressions in the classical representation is provided in \cref{app:dictionary}.
\end{enumerate}
 In all three approaches, we start with a simplified setting containing only one molecular species, which drastically simplifies the notation, and then generalize to reactions involving several species such as, complex formation and general association reactions.

\section{The chemical diffusion master equation: an intuitive formulation} \label{sec:CDME}

We consider an open system of a varying number of diffusing particles of the same chemical species in a finite space domain $\mathbb{X}\subset \mathbb{R}^d$. The diffusion process changes the spatial configuration of the particles while the reaction process can change the number of particles in the system. The configuration of the system is thus given by the numbers of particles and their positions. The probability distribution of such a system is given as an ordered family of probability density functions:
\begin{align}
	\rho = \left( \rho_0, \rho_1, \rho_2, \dots, \rho_n, \dots \right),
\end{align}

where $\rho_n(x^{(n)})$ is the probability density of finding $n$ particles at the positions $x^{(n)}=(x_1^{(n)},\dots, x_n^{(n)})$ for $n\geq 1$, while $\rho_0$ is the probability for no particles being present. As the particles are statistically indistinguishable from each other, the densities must be symmetric with respect to permutations of particle labels, e.g. $\rho_2(y,z) = \rho_2(z,y)$ for all $y,z \in \mathbb{X}$, and more generally
\begin{equation} \rho_n(x^{(n)}) = \rho_n(Px^{(n)}) \quad \text{for all} \: P \in \mathcal{P}_n  \end{equation}
where $\mathcal{P}_n$ is the set of all permutations of an $n$-tuple. 
%\fh{Please remove the brackets after the permutation operators below.}
The normalization condition is 
\begin{align} \label{eq:normalization}
	\rho_0 + \sum_{n=1}^\infty \int_{\mathbb{X}^n} \rho_n(x^{(n)})dx^{(n)} = 1.
\end{align}
In general, $\rho$ will also depend on time, $\rho_n=\rho_n(t,x^{(n)})$, but we will omit $t$ for simplicity. As a remark, the distribution $\rho$ is an element of a linear function space similar to the Fock space of quantum mechanics, see refs.~\onlinecite{del2021probabilistic, doi1976second, grassberger1980fock} and \cref{sec:FockCDME}.

Given that there are $M\in \mathbb{N}$ reactions, the CDME has the general form
\begin{equation}\label{CDME}
	\frac{\partial \rho}{\partial t} = \left(\mathcal{D} + \sum_{r=1}^M\mathcal{R}^{(r)} \right)\rho
\end{equation}
for a diffusion operator $\mathcal{D}$ and reaction operators $\mathcal{R}^{(r)}$. 
Each of the reaction operators $\mathcal{R}^{(r)}$ corresponds to one possible reaction, and it is conveniently split into loss and gain operators\footnote{Similarly, the reaction operator in ref.~\onlinecite{del2021probabilistic} was split into a particle conserving part (the loss operator) and a non-conserving part (the gain operator).},
\begin{align}
	\mathcal{R}^{(r)} = \mathcal{G}^{(r)} - \mathcal{L}^{(r)}. 
\end{align}
In the following, we will construct these loss and gain operators at first for reactions of a single species and then for a multi-species scenario. In each case, we consider a system with only one reaction, such that the index $r$ can be skipped. For systems with several reactions, the results may simply be combined by summing up these operators as in \cref{CDME}.

\subsection{One species}\label{CMDE_onespecies}

To start with, we assume that there is only one chemical species~$A$. The most general reaction in this case is of the form
\begin{equation}
kA\rightarrow lA
\end{equation}
for $k,l \in \mathbb{N}_0$. 
The rate at which a reaction event occurs is given by $\lambda(y^{(l)};x^{(k)})>0$, and it depends on the positions $x^{(k)}\in \mathbb{X}^k$ of the reactants and the positions $y^{(l)}\in \mathbb{X}^l$ of the products. Note that the rate function $\lambda$ should be symmetric with respect to pair exchanges in both of its arguments. 

We can now write the $n$th component of the CDME as
\begin{equation}\label{CDME_n}
	\frac{\partial\rho_n}{\partial t} = \mathcal{D}_n \rho_n + \mathcal{G}_{n} \rho_{n+k-l} - \mathcal{L}_n \rho_n
\end{equation}
for appropriate operators $\mathcal{D}_n$, $\mathcal{G}_{n}$, $\mathcal{L}_n$ referring to diffusion, gain and loss, respectively\footnote{In ref.~\onlinecite{del2021probabilistic}, the loss operator was denoted by $\mathcal{R}^{(k)}$, and the gain operator as $\mathcal{R}^{(k,l)}$. We find the new notation less cumbersome.}.
Reactions at the $n$-particle state produce a transition to the $(n-k+l)$-particle state. Thus, the loss of probability for the $n$-particle state $\rho_n$ depends only on itself. Similarly, reactions at the $(n+k-l)$-particle state produce a transition to the $\rho_n$ state. Thus, the gain of probability for the $n$-particle state depends on $\rho_{n+k-l}$.

For physically non-interacting particles, the diffusion operator $\mathcal{D}_n$ can be expressed in terms of the one-particle diffusion $D_\nu$ applied to the $\nu$th particle:
\begin{align}
	\mathcal{D}_n &= \sum_{\nu=1}^n D_\nu, 
	\label{eq:diffope}
\end{align}
where $D_\nu$ is the infinitesimal generator of the one-particle Fokker--Planck equation.
For example, one may think of $D_\nu$ as something as simple as the $d$-dimensional Laplacian, $D_\nu=\nabla^2_{x_\nu}$. Ignoring the reaction operators and assuming that there is no exchange of particles with a reservoir outside of $\mathbb{X}$ \cite{klein2022derivation}, all the resulting equations are uncoupled and one obtains a family of uncoupled Fokker--Planck equations, unless there is an exchange of particles with the world outside of $\mathbb{X}$, in which case one ends up again with a similar family of many-particle densities, albeit with a different structure of the coupling
between its levels \cite{delle2020liouville}. For simplicity of the exposition, we assume reflecting boundaries for $\mathbb{X}$ from here on, i.e., a confinement by rigid walls.

\begin{figure*}
	\includegraphics[width=\textwidth]{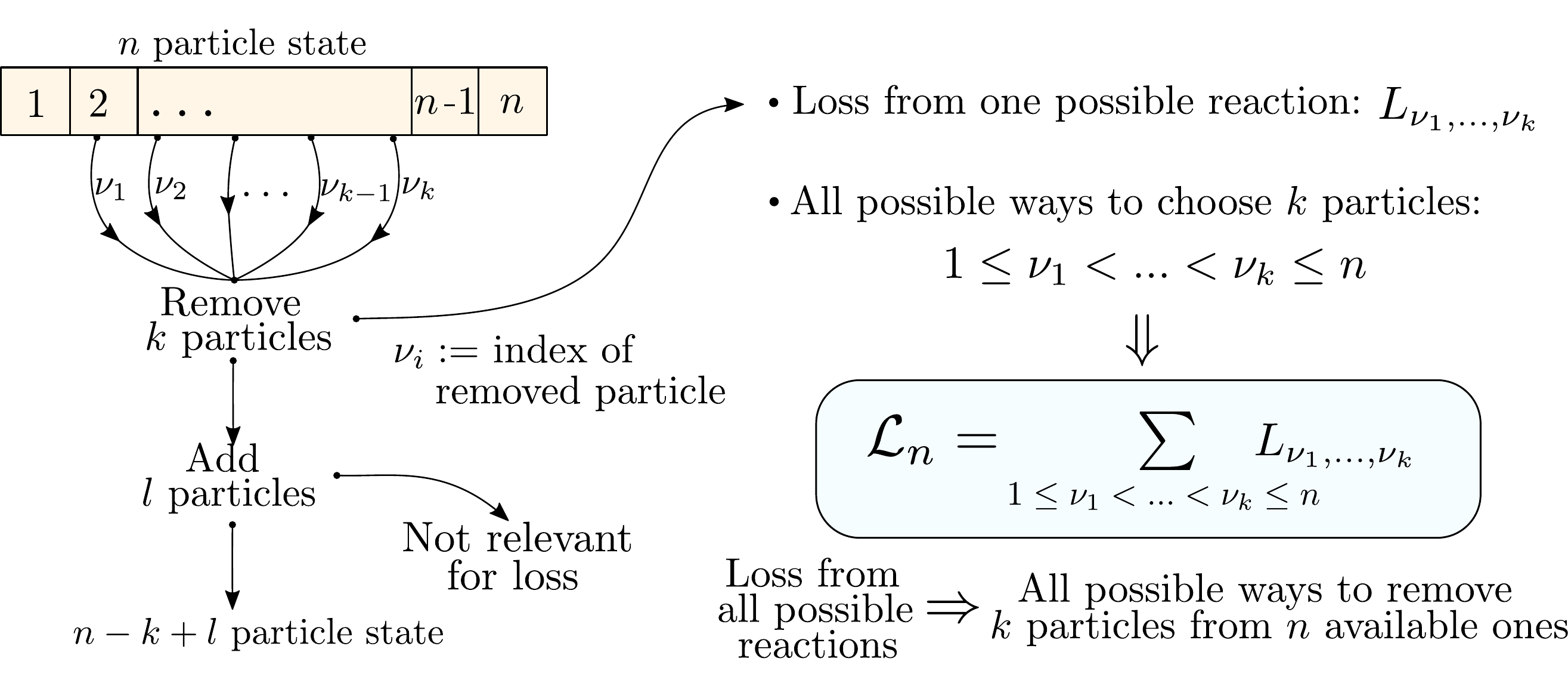}
	\caption{Diagram representing the loss of probability from the $n$-particle state due to the reaction $kA\rightarrow lA$ (\cref{eq:losskA-lA}). The particle states are represented by a set of boxes, where each box correspond to the index of a particle.}
	\label{fig:loss}
\end{figure*}

The loss operator acting on the $n$-particle density will output the total rate of probability loss of $\rho_n$ due to all possible combinations of reactants. It is given in terms of the loss per reaction $L_{\nu_1,\dots, \nu_k}$ (\textit{local} loss), which acts on $k$ particles at a time, with $(\nu_1,\dots, \nu_k)$ denoting the indexes of the particles that it acts on. The loss per reaction quantifies how much probability is lost to the current state due to one reaction, it is thus the integral over the density and the rate function $\lambda$ over all the possible positions of the products:
\begin{align}
\left( L_{\nu_1,\dots, \nu_k} \rho_n \right) (x^{(n)}) = \rho_n(x^{(n)}) \int_{\mathbb{X}^l} \lambda(y^{(l)};x^{(n)}_{\nu_1,\dots,\nu_k}) dy^{(l)},
\label{eq:lossperreaction}
\end{align}
where $x^{(n)}_{\nu_1,\dots,\nu_k} := (x^{(n)}_{\nu_1}, \dots, x^{(n)}_{\nu_k})$.
The total loss is then the sum of the loss per reaction over all possible reactions,
\begin{align}
	\mathcal{L}_n= \sum_{1\leq \nu_1 < \dots <\nu_k \leq n}L_{\nu_1,\dots, \nu_k}.
	\label{eq:losskA-lA}
\end{align}
The form of the ordered sum guarantees that we count all the possible ways of picking up $k$ particles without double counting, see \cref{fig:loss} for a diagram of the calculation. For the special case of $k=0$ we have
\begin{equation} (\mathcal{L}_n \rho_n)(x^{(n)})=  \rho_n(x^{(n)}) \int_{\mathbb{X}^l} \lambda(y^{(l)}) dy^{(l)}.  \end{equation}

\begin{figure*}
	\includegraphics[width=\textwidth]{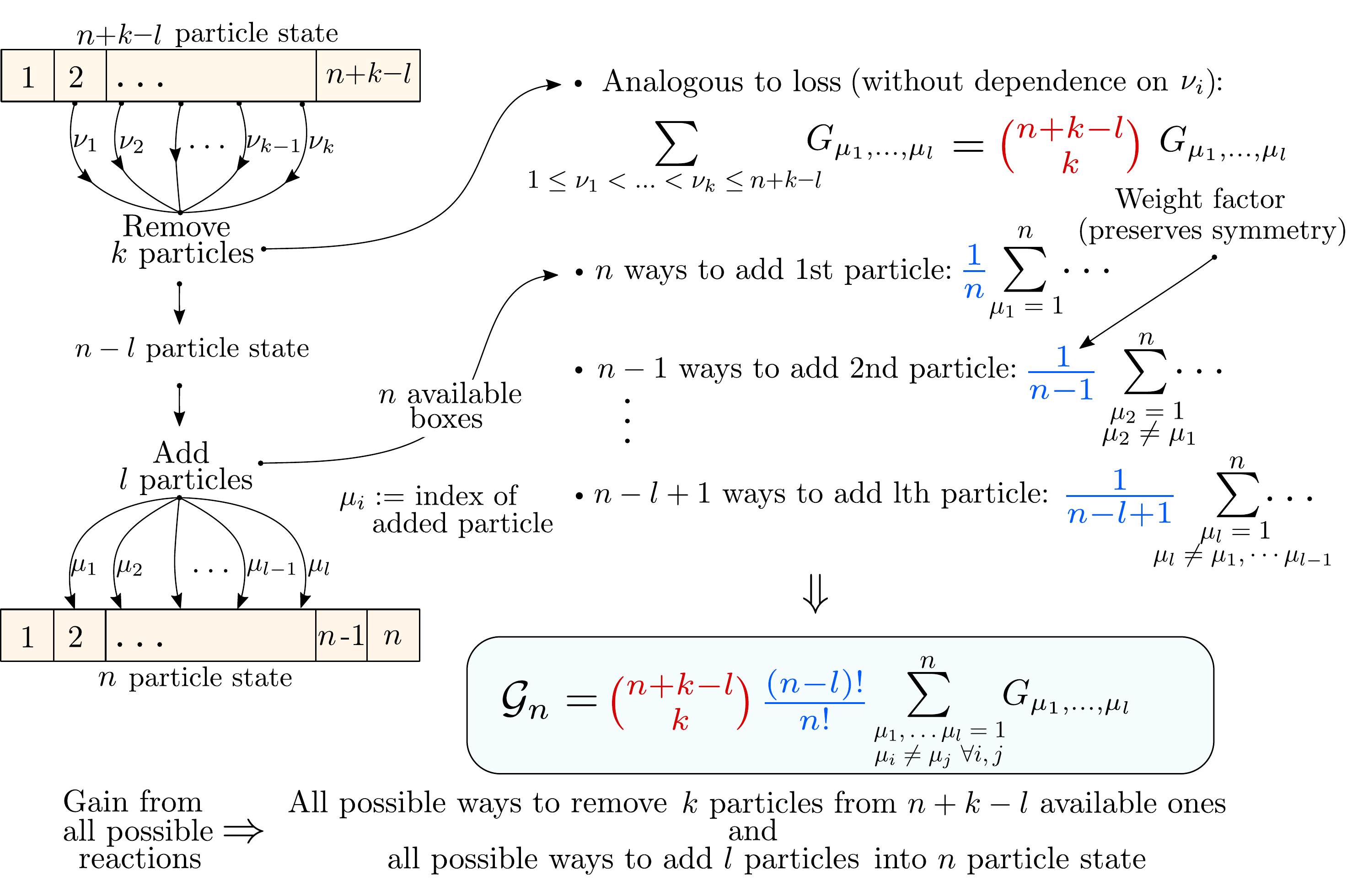}
	\caption{Diagram representing the gain of probability for the $n$-particle state for the reaction $kA\rightarrow lA$ (\cref{eq:gainkA-lA}). The particle states are represented by a set of boxes, where each box correspond to the index of a particle. The final expression can be further simplified, see \cref{eq:gainkA-lA2}.
	}
	\label{fig:gain}
\end{figure*}

Similarly, the gain operator acting on the $n$-particle density will output the total rate of probability gain of $\rho_n$. It can be expressed in terms of the gain per reaction (\textit{local} gain) resulting from  $k$ reacting particles with indexes $(\nu_1,\dots, \nu_k)$ producing $l$ products with indexes $(\mu_1,\dots, \mu_l)$, termed $G_{ \mu_1,\dots,\mu_l}$. The gain per reaction quantifies how much probability is gained by the current state due to one reaction, it is thus the integral over the density and the rate function $\lambda$ over all the possible positions of the reactants:
\begin{align}
	\left(G_{\mu_1,\dots,\mu_l} \rho_{n+k-l} \right) (x^{(n)}) = \int_{\mathbb{X}^k}
    \lambda(x^{(n)}_{\mu_1,\dots,\mu_l};z^{(k)})
    \rho_{n+k-l}(x^{(n)}_{\setminus\{\mu_1,\dots,\mu_l \}}, z^{(k)})  dz^{(k)},
	\label{eq:gainperreaction}
\end{align}

where the subscript $\setminus \{\mu_1,\dots,\mu_l\}$ means that the entries with indices $\mu_1, \dots, \mu_l$ are excluded from the tuple $x^{(n)}$ of particle positions. Note that the indexes of the reacting particles $\nu_1,\dots,\nu_k$ are not relevant for the gain since the reactants' positions are integrated out (and both the density and the rate function are symmetric). The total gain is then the sum of the gain per reaction over all possible reactions,

\begin{subequations} \label{eq:gainkA-lA}
\begin{align}
\mathcal{G}_n & =  \frac{(n-l)!}{n!} \binom{n+k-l}{k} \ \
\sum_{\substack{\mu_1\dots\mu_l=1 \\ \mu_i \neq \mu_j \ \forall i,j}}^{n} G_{\mu_1,\dots,\mu_l}  \label{eq:gainkA-lA1} \\
& = \binom{n}{l}^{-1} \binom{n+k-l}{k} \ \
\sum_{1\leq\mu_1<\dots<\mu_l\leq n}   G_{\mu_1,\dots,\mu_l}, \label{eq:gainkA-lA2}
\end{align}
\end{subequations}
where we used the symmetry of $G_{\mu_1,\dots,\mu_l}$ with respect to the indices.
The complicated form of the gain operator is due to the fact that it needs to consider all the possible ways to pick up $k$ particles from the $n+k-l$-particle state, just as the loss operator, but in addition, it also needs to consider all the possible ways of incorporating $l$ particles into the current state in a symmetry-preserving manner, see \cref{fig:gain} for a diagram illustrating the calculation.
Note that the output of the loss and gain operators is also symmetric.

Let us use the preceding formulas for general reactions involving one species to derive the CDME for some common reactions (for simplicity, we write $\rho_n(x^{n},t)$ as $\rho_n(x^{n})$):
\begin{itemize}
	\item Degradation $A\rightarrow \emptyset$:
	This case is recovered with $k=1,l=0$ using the rate function $\lambda_d(x) = \lambda_d(;x)$. The CDME reads
	\begin{align}\label{CDME:decay}
		\frac{\partial\rho_n}{\partial t}(x^{n})
		&= \sum_{\nu=1}^n D_\nu \rho_n(x^{n}) + (n+1)\int_{\mathbb{X}} \lambda_d(z)\rho_{n+1}(x^{(n)},z) dz -\rho_n(x^{(n)}) \sum_{\nu=1}^{n}   \lambda_d(x^{(n)}_\nu).
	\end{align}
	\item Creation $\emptyset \rightarrow A$:
	Here, we set $k=0$, $l=1$ using the rate function $\lambda_c(y) = \lambda_c(y;)$, then the CDME is
	\begin{align}
		\frac{\partial\rho_n}{\partial t}(x^{n}) &= \sum_{\nu=1}^n D_\nu \rho_n(x^{n})  + \frac{1}{n}\sum_{\mu=1}^{n}\rho_{n-1}\bigl(x^{(n)}_{\setminus\{\mu\}}\bigr)  \lambda_c(x^{(n)}_\mu) - \rho_n(x^{(n)}) \int_{\mathbb{X}} \lambda_c(y) dy.
	\end{align}
	\item Mutual annihilation $A + A \rightarrow \emptyset$: In this case, we have $k=2$, $l=0$ with the rate function $\lambda_a(x_1,x_2)=\lambda_a(;x_1,x_2)$. Then
	\begin{multline}
		\frac{\partial\rho_n}{\partial t}(x^{n})
		= \sum_{\nu=1}^n D_\nu \rho_n(x^{n})  + \frac{(n+2)(n+1)}{2}\int_{\mathbb{X}^2} \lambda_a(z_1,z_2)\rho_{n+2}(x^{(n)},z_1,z_2) dz_1dz_2 \\
		-\rho_n(x^{(n)}) \sum_{1\leq\nu_1<\nu_2\leq n}   \lambda_a(x^{(n)}_{\nu_1},x^{(n)}_{\nu_2}).
	\end{multline}
	\item Trimolecular reaction: $3A \rightarrow 2A$: Here, $k=3$, $l=2$ and the rate function is $\lambda(y^{(2)};x^{(3)})$, then
	\begin{multline}
		\frac{\partial\rho_n}{\partial t}(x^{n})
		=  \sum_{\nu=1}^n D_\nu \rho_n(x^{n}) + %	 \frac{(n-2)!}{n!}\frac{(n+1)!}{(n-2)!3!}
		\frac{n+1}{3}
		\sum_{1\leq\mu_1<\mu_2\leq n}\int_{\mathbb{X}^3} \lambda(y^{(2)};z^{(3)})\rho_{n+1}(x^{(n)}_{\setminus \{\mu_1, \mu_2\} },z^{(3)}) dz^{(3)}  \\
		  -  \rho_n(x^{(n)})\sum_{1\leq\nu_1<\nu_2<\nu_3\leq n} \int_{\mathbb{X}^2}\lambda(y^{(2)}; x^{(n)}_{\nu_1},x^{(n)}_{\nu_2},x^{(n)}_{\nu_3})dy^{(2)}.
	\end{multline}
\end{itemize}

\paragraph*{Several reactions:} Given a system with several reactions of the form $k_rA \to l_rA$ for different $k_r,l_r \in \mathbb{N}_0$, $r=1,\dots,M$, the $n$th component of the CDME is given by a sum of the form
\begin{equation}\label{CDME_n_sum}
\frac{\partial\rho_n}{\partial t} = \mathcal{D}_n \rho_n + \sum_{r=1}^M \left( \mathcal{G}^{(r)}_{n} \rho_{n+k_r-l_r} - \mathcal{L}^{(r)}_n \rho_n \right)
\end{equation}
with accordingly defined operators $\mathcal{G}^{(r)}_{n}$ and $ \mathcal{L}^{(r)}_n$. 

As one can see from the expressions above, the explicit formulation of the loss and gain operators can become quite complex due to the combinatorics. This issue worsens when several species are involved. Thus, it appears convenient to have a formalism where the combinatorial factors are intrinsically built-in \cite{del2021probabilistic}, and we will present such an approach in \cref{sec:FockCDME}. Beforehand, we will explore one example with multiple species, as well as an alternative explicit representation of the CDME.

\subsection{Multiple species}\label{sec:ABC}

Consider the reaction 
\begin{equation}
A+B \rightarrow C
\end{equation}
with rate function $\lambda(y; x_A,x_B)$, where $x_A$ and $x_B$ are the locations of one pair of reactants and $y$ is the location of the product. The stochastic dynamics of the system is described in terms of the distributions $\rho_{a,b,c}\left(x^{(a)},x^{(b)},x^{(c)}\right)$, where $a,b,c$ indicate the numbers of $A$, $B$, and $C$ particles, respectively, and $x^{(a)}$ indicates the positions of the $A$ particles, $x^{(b)}$ of the $B$ particles, and $x^{(c)}$ of the $C$ particles.\footnote{When dealing with systems with one species, we will use $n$ to specify the number of particles. Otherwise, we denote the number of particles by the lower case letter of the corresponding species.} The normalization condition \cref{eq:normalization} generalizes to
\begin{align}
	\sum_{a,b,c=0}^\infty\; \int\limits_{\mathbb{X}^a\times\mathbb{X}^b\times\mathbb{X}^c} \hspace{-0.5cm} \rho_{a,b,c}\left(x^{(a)},x^{(b)},x^{(c)}\right) dx^{(a)}\,dx^{(b)}\,dx^{(c)} = 1.
\end{align}
The CDME for this reaction has the same structure as before, namely $\partial \rho/\partial t = \mathcal{D}\rho + \mathcal{R} \rho$ for a diffusion operator $\mathcal{D}$ and a reaction operator $\mathcal{R}$. Writing the equation component-wise and separating the reaction operator into its total loss and gain operators, we obtain
\begin{align}
	\frac{\partial \rho_{a,b,c}}{\partial t} = \mathcal{D}\rho_{a,b,c} + \mathcal{G}_{a,b,c} \rho_{a+1,b+1,c-1} - \mathcal{L}_{a,b} \rho_{a,b,c}.
\end{align}
The total loss and gain operators can be written explicitly by defining them per reaction (\textit{locally}) and applying them to all possible combinations of reactors and products in the corresponding state (\textit{globally}). Following the same logic as in \cref{fig:loss}, the loss operator is given by
\begin{equation}
	\mathcal{L}_{a,b} = \sum_{\mu=1}^a \sum_{\nu=1}^b L_{\mu,\nu}
\end{equation}
with
\begin{equation}
	\left(L_{\mu,\nu} \ \rho_{a,b,c}\right)(x^{(a)},x^{(b)},x^{(c)})  = \rho_{a,b,c}\left(x^{(a)},x^{(b)},x^{(c)}\right) \int_{\mathbb{X}} \lambda\left(y; x_{\mu}^{(a)},x_{\nu}^{(b)}\right) dy.
	%	\left(x^{(a)},x^{(b)},x^{(c)}\right)
\end{equation}
Note that for the loss the positions of the products are not relevant, so $L_{\mu,\nu}$ just depends on the indexes $\mu,\nu$ of the reactants. Moreover, in contrast to \cref{eq:losskA-lA}, the sum is not ordered since the reaction involves different species. Analogously, we can write the gain, but it is usually more complex since now the location of the products do matter.
In analogy to \cref{fig:gain}, the gain operator is: 
\begin{equation}
 	\mathcal{G}_{a,b,c} =\frac{1}{c}\sum_{\xi=1}^{c}\sum_{\mu=1}^{a+1} \sum_{\nu=1}^{b+1} G_{\xi} = (a+1) (b+1)\frac{1}{c}\sum_{\xi=1}^{c} G_{\xi}
\end{equation}
with
\begin{multline}
	\left( G_{\xi} \rho_{a+1,b+1,c-1}\right)(x^{(a)},x^{(b)},x^{(c)}) =
    \\ \int_{\mathbb{X}^2}\lambda\left(x^{(c)}_\xi; z,z'\right) \rho_{a+1,b+1,c-1}
	\left( (x^{(a)},z),(x^{(b)},z),x^{(c)}_{\setminus \{\xi\}} \right)
	dz\,dz'.
	%	\left(x^{(a)},x^{(b)},x^{(c)}\right)
\end{multline}
Gathering the terms and incorporating the diffusion term in the same way as before for each species, we obtain the CDME
\begin{align}\begin{split}\label{CDME:ABC}
		\frac{\partial \rho_{a,b,c}}{\partial t} = & \sum_{\mu=1}^a D^A_{\mu}\rho_{a,b,c} + \sum_{\nu=1}^b D^B_{\nu}\rho_{a,b,c} +\sum_{\xi=1}^c D^C_{\xi}\rho_{a,b,c}  \\
				& +\frac{(a+1)(b+1)}{c}\sum_{\xi=1}^{c}  \int_{\mathbb{X}^2}	\lambda\left(x^{(c)}_\xi; z,z'\right) \rho_{a+1,b+1,c-1}
		\left( (x^{(a)},z),(x^{(b)},z'),x^{(c)}_{\setminus \{\xi\}} \right)
		dz\,dz' \\
		&- \rho_{a,b,c}\left(x^{(a)},x^{(b)},x^{(c)}\right)\sum_{\mu=1}^a \sum_{\nu=1}^b  \int_{\mathbb{X}} \lambda\left(y; x_{\mu}^{(a)},x_{\nu}^{(b)}\right) dy, \\
	\end{split}
\end{align}
where in the dependence of $\rho_{a,b,c}$ on the positions $\left(x^{(a)},x^{(b)},x^{(c)}\right)$ and time $t$ has been skipped in the first line to simplify notation.

We see again that the main difficulty in writing down the CDME correctly is to come up with expressions that relate the loss and gain operators acting on a subset of particles to the loss and gain operators acting on the whole system. This is expected as the operators need to account for all possible combinations of particles that can undergo a certain reaction.

\section{CDME formulation using many-particle propensities} \label{sec:ManyParticlePropensity}

In this section, we provide another justification of the form of the gain and loss operators (especially of the combinatorial factors) by utilizing permutations and Dirac $\delta$-distributions to mathematically describe the particle selection process, and by transforming the local rate function into many-particle propensity functions.

For the simplicity of the notation, we again restrict to the case of only one chemical species as in \cref{CMDE_onespecies}; a case with multiple species will be discussed in \cref{sec:prop_multiplespecies}. Given the component-wise formulation \cref{CDME_n_sum} of the CDME, we would like to express the gain and loss operators by means of global \textit{many-particle propensities}, which express the likeliness for a reaction to take place depending on the whole system state.
More concretely, given a single reaction of the form $kA\to lA$, we consider for each $n$ the propensity functions $\Lambda_n: \mathbb{X}^{n-k+l} \times \mathbb{X}^n \to [0,\infty) $, where $\Lambda_n(y^{(n-k+l)};x^{(n)})$ refers to the probability per unit of time that a system with $n$ particles in the ordered positions $x^{(n)}_1,\dots,x^{(n)}_n$ gets to be transformed into a system with $n-k+l$ particles in the ordered positions $y^{(n-k+l)}_1,\dots,y^{(n-k+l)}_{n-k+l}$.

Reflecting the assumption that particles of a single species are modeled as indistinguishable, the many-particle propensities are required to be symmetric with respect to pair exchanges in both of their arguments. 

In terms of the many-particle propensities $\Lambda_n$, the loss and gain operators are given by
\begin{subequations}\label{loss-gain}\begin{align}
	(\mathcal{L}_n\rho_n)(x^{(n)}) & = \rho_n(x^{(n)}) \int_{\mathbb{X}^{n-k+l}} \Lambda_n(y^{(n-k+l)};x^{(n)}) dy^{(n-k+l)}, \label{eq:LDefinition}\\
	(\mathcal{G}_n\rho_{n+k-l})(y^{(n)}) & = \int_{\mathbb{X}^{n+k-l}} \Lambda_{n+k-l}(y^{(n)};x^{(n+k-l)}) \rho_{n+k-l}(x^{(n+k-l)}) dx^{(n+k-l)}, \label{eq:GDefinition}
	\end{align}
\end{subequations}

in analogy to the operators given in \cref{CMDE_onespecies}.  These expressions are symmetry preserving owing to the symmetry properties of the densities and of the propensities. They are probability preserving, too, because taking into account that $\mathcal{L}_n\rho_n$ is a loss for $\rho_n$ while $\mathcal{G}_{n-k+l}\rho_{n}$ is a gain for $\rho_{n-k+l}$, the sum of the changes of total probability in the $n$- and $n-k+l$-particle spaces due to the considered reaction is
\begin{equation}\label{loss=gain}
\int_{\mathbb{X}^n}  (\mathcal{L}_n\rho_n)(x^{(n)}) dx^{(n)} - \int_{\mathbb{X}^{n-k+l}} (\mathcal{G}_{n-k+l}\rho_{n})(y^{(n-k+l)}) dy^{(n-k+l)}=0.
\end{equation}
If the densities $\rho_n$ are symmetric with respect to arbitrary particle permutations initially, the loss and gain operations from \cref{eq:LDefinition,eq:GDefinition} will preserve this property. Moreover, owing to the way the densities are normalized in \cref{eq:normalization}, no normalizing combinatorial factors arise in \cref{loss-gain,loss=gain};
rather, the combinatorics is hidden in the definition of $\Lambda_n$.
Thus, we conclude that preservation of symmetry and probability is straightforwardly ensured when working with the many-particle propensities $\Lambda_n$.

Given a finite number $M$ of reactions of the form $k_rA \to l_rA$, we denote the propensity functions of the $r$th reaction by $\Lambda^{(r)}_n$ and the corresponding loss and gain operators by $\mathcal{L}^{(r)}_n$ and $\mathcal{G}^{(r)}_n$. Inserting into \cref{CDME_n_sum}, we obtain the $n$th component of the CDME in terms of the many-particle propensities  $\Lambda^{(r)}_n$. 

The next step is to derive the concrete form of the many-particle propensity $\Lambda_n$ for specific reactions and relate them to the local rate functions $\lambda$. Remember that, in contrast to the propensities $\Lambda_n$, the rate functions $\lambda$ define the rate for a reaction taking place solely depending on the positions of reactants and products. More concretely, $\lambda(y^{(l)},x^{(k)})$ defines the probability per unit of time for $k$ particles located at $x^{(k)}_1,\dots,x^{(k)}_k$ to be fully replaced due to the reaction $kA \to lA$ by $l$ particles located at $y^{(l)}_1,\dots,y^{(l)}_l$. In contrast, the global many-particle propensities $\Lambda_n$ depend on the complete system state before and after the reactions and already contain combinatorial factors and symmetrization. As a first scenario, we consider the example of simply decay.

\subsection{Many-particle propensity for simple decay}

Here we develop an explicit formula that relates the reaction rate $\lambda_d(x)$ of the decay process, see \cref{CDME:decay}, to the associated many-particle propensity $ \Lambda_{n+1}:\mathbb{X}^{n} \times \mathbb{X}^{n+1} \to [0,\infty)$. 
The following formula captures the essence of the remaining many-particle propensities but does not yet respect the required symmetries and the normalization,

\begin{equation}\label{eq:PropensityDecayBeforeSymmetrization}
	\Lambda_{n+1}^{\text{bs}}\bigl(y^{(n)},x^{(n+1)}\bigr) 
= \sum_{\nu=1}^{n+1} \lambda_d\bigl(x_\nu^{(n+1)}\bigr) \delta^n\mleft(x^{(n+1)}_{\setminus \{\nu \}}- y^{(n)}\mright)
\end{equation}
where the superscript ``bs'' stands for ``before symmetrization'', and $\delta^n$ refers to the Dirac distribution in $n$ dimensions; in particular,
\begin{equation}
 \delta^n\mleft(x^{(n+1)}_{\setminus \{\nu \}}-y^{(n)}\mright)
  = \prod_{\mu=1}^{\nu-1} \delta\mleft(x^{(n+1)}_{\mu\vphantom{+1}}-y^{(n)}_\mu\mright)
    \prod_{\mu=\nu}^n \delta\mleft(x^{(n+1)}_{\mu+1}-y^{(n)}_\mu\mright).
\end{equation}
The term under the sum in \cref{eq:PropensityDecayBeforeSymmetrization} describes (i) the probability per unit time that the $\nu$th particle disappears from position $x\ord{n+1}_\nu$, and (ii) the fact that the rest of the configuration remains unchanged, so that its probability is transferred from $\rho_{n+1}\bigl(x\ord{n+1}\bigr)$ to~$\rho_n\bigl(x\ord{n+1}_1, \dots, x\ord{n+1}_{\nu-1}, x\ord{n+1}_{\nu+1}\dots, x\ord{n+1}_{n+1}\bigr)$. The summation over $\nu$ accounts for the fact that any of the particles out of configuration $x^{(n+1)}$ might decay.

The properly symmetrized version of \cref{eq:PropensityDecayBeforeSymmetrization} is obtained by averaging over all permutations of the target space configurations $y^{(n)}$, i.e.,
\begin{equation}\label{eq:PropensityDecay}
	\Lambda_{n+1}\bigl(y^{(n)},x^{(n+1)}\bigr) 
	= \frac{1}{n!} \sum_{P\in\mathcal{P}_n}  
	\Lambda_{n+1}^{\text{bs}}\left(P y^{(n)},x^{(n+1)}\right)\,.
\end{equation}
Owing to the summation over $\nu$ in \cref{eq:PropensityDecayBeforeSymmetrization}, this formula is already symmetric with respect to permutations of the second argument $x^{(n+1)}$. In turn, averaging over the permutations in $\mathcal{P}_n$ guarantees that the probability associated with a particle disappearing from the $(n+1)$-particle configuration $x^{(n+1)}$ is distributed symmetrically to that of all equivalent $n$-particle configurations on the receiving end.

Now, the crucial step is to insert the propensities into \cref{loss-gain} and translate the expressions given in \cref{eq:PropensityDecay} into combinatorial factors.
Due to the particle exchange symmetry of $\rho_{n+1}$, the contribution of any of the terms under the sum in \cref{eq:PropensityDecay} to $(\mathcal{G}_n\rho_{n+1})\bigl(y^{(n)}\bigr)$ from \cref{eq:GDefinition} obeys (see also \cref{eq:PropensityDecayBeforeSymmetrization})
\begin{equation}\label{eq:PropensityDecayGEvaluation}
	\begin{array}{l}
		\dss \int_{\mathbb{X}^{n+1}} 
		 \lambda_d\left(x^{(n+1)}_\nu\right)
	\delta^n\left(x^{(n+1)}_{\setminus \{\nu\}} - P y^{(n)}\right)
		\rho_{n+1}\bigl(x^{(n+1)}\bigr) dx^{(n+1)}
		\\[10pt]
		\dss 
		= \int_{\mathbb{X}} 
		\lambda_d (x) \
		\rho_{n+1}\!\left(\bigl(P y^{(n)}\bigr)_1, \dots, \bigl(P y^{(n)}\bigr)_{\nu-1}, x, \bigl(P y^{(n)}\bigr)_{\nu}, \dots,  \bigl(P y^{(n)}\bigr)_n \right) dx
		\\[10pt]
		\dss 
		= \int_{\mathbb{X}} 
		\lambda_d (x)\ 
		\rho_{n+1}\left(P y^{(n)}, x\right)\, dx
		= \int_{\mathbb{X}} 
		\lambda_d (x)\ 
		\rho_{n+1}\bigl(y^{(n)}, x\bigr)\, dx\,,
	\end{array}
\end{equation}
i.e., they are all the same. Summation of this expression over $\nu$ (see \cref{eq:PropensityDecayBeforeSymmetrization}) yields a factor of $(n+1)$ and summation over the $n$-particle permutations $P\in \mathcal{P}_n$ together with the division by $n!$ (see \cref{eq:PropensityDecay}) ensures that the $(n+1)$-particle probability is distributed symmetrically over the $n$-particle space.

A similar calculation for the loss $\left(\mathcal{L}_n\rho_n\right)(x^{(n)})$ reads 
\begin{equation}\label{eq:PropensityDecayLEvaluation}
		\rho_{n}\bigl(x^{(n)}\bigr) \lambda_d\left(x^{(n)}_\nu\right)
		\int_{\mathbb{X}^{n-1}} 
		\delta^{n-1}\left(x^{(n)}_{\setminus \{\nu \}}- Py^{(n-1)}\right) 
		dy^{(n-1)}
		= \rho_{n}\bigl(x^{(n)}\bigr) \lambda_d\left(x^{(n)}_\nu\right)
\end{equation}
for each permutation $P$ and each index $\nu$, where we translated formula  \cref{eq:PropensityDecay} for $\Lambda_{n+1}$ to  $\Lambda_n$ by a shift in $n$. Summation over the $n$-particle permutations $P\in \mathcal{P}_n$ cancels the factor $1/n!$.
Summation over $\nu$, and combining with the result for the gain, we obtain the evolution equation for the $n$-particle density under a simple decay process:
\begin{equation}
	\partial_t \rho_{n}\bigl(x^{(n)}\bigr) 
	= (n+1) \int_{\mathbb{X}} \lambda_d (y)\ \rho_{n+1}\bigl(x^{(n)}, y\bigr)\, dy
	- \rho_{n}\bigl(x^{(n)}\bigr) \ \sum_{\nu=1}^{n} \lambda_d\left(x^{(n)}_\nu\right)\,,
\end{equation}
and this is in line with the reaction terms in \cref{CDME:decay}.

\subsection{Many-particle propensity for multiple species}\label{sec:prop_multiplespecies}

We continue with the scenario of multiple species as described in \cref{sec:ABC}. Let $\lambda(y;x_A,x_B)$ again denote the conditional probability per unit time that the reaction $A+B\to C$ occurs with a product particle of species $C$ appearing in $y$, given that two reactants $A$ and $B$ reside in $x_A$ and $x_B$, respectively. Then we are interested in the associated many-particle propensities

\begin{equation}
	\Lambda\left(y^{(a-1)},y^{(b-1)},y^{(c+1)}; x^{(a)},x^{(b)},x^{(c)}\right),
\end{equation}
which denotes the transfer of probability density per unit time from $\rho_{a,b,c}$ to $\rho_{a-1,b-1,c+1}$ due to the considered reaction. Note that we have here suppressed the subscript $a,b,c$ on $\Lambda$ to simplify notation. 

At first, we define for each tuple of indices $\nu,\mu,\xi$ the propensity
\begin{equation}\label{eq:lambdaBareIndividual_ABC}
\begin{array}{ll}
\dss \Lambda_{\nu,\mu,\xi}\left(y^{(a-1)},y^{(b-1)},y^{(c+1)}; x^{(a)},x^{(b)},x^{(c)}\right) \span\omit
\\[10pt]
= & \dss \lambda\bigl(y^{(c+1)}_\xi; x^{(a)}_\mu, x^{(b)}_\nu\bigr) 
	\delta^{a-1} \left(x^{(a)}_{\setminus \{\mu \}} - y^{(a-1)} \right)
	\delta^{b-1} \left(x^{(b)}_{\setminus \{\nu \}} - y^{(b-1)} \right)
	\delta^c \left(x^{(c)} - y^{(c+1)}_{\setminus \{\xi \}} \right).
\end{array}
\end{equation}

The interpretation of the expression in \cref{eq:lambdaBareIndividual_ABC} is as follows: Given the reactant and product tuples in the source and target spaces, $\bigl(x^{(a)},x^{(b)},x^{(c)}\bigr)$ and $\bigl(y^{(a-1)},y^{(b-1)},y^{(c+1)}\bigr)$, respectively, it assigns the (probability) transfer rate $\lambda(y^{(c+1)}_\xi;x^{(a)}_\mu, x^{(b)}_\nu)$ to the reaction occurring between the reactants located at $x^{(a)}_\mu, x^{(b)}_\nu$ and producing a product particle in $y^{(c+1)}_\xi$. The products of $\delta$-distributions make sure that in the transfer all other particle positions remain those from the source space tuples.

In analogy to \cref{eq:PropensityDecayBeforeSymmetrization}, we can now write down the many particle propensity before symmetrization as

\begin{multline}\label{eq:lambdaBare_ABC}
\Lambda^{\text{bs}}\left(y^{(a-1)},y^{(b-1)},y^{(c+1)}; x^{(a)},x^{(b)},x^{(c)}\right)
= \\ \frac{1}{c+1} \sum_{\mu=1}^{a}\sum_{\nu=1}^{b} \sum_{\xi=1}^{c+1} \Lambda_{\nu,\mu,\xi}\left(y^{(a-1)},y^{(b-1)},y^{(c+1)}; x^{(a)},x^{(b)},x^{(c)}\right).
\end{multline}
The prefactor of $1/(c+1)$ is to be included for the following reason: If
$y^{(a-1)},y^{(b-1)}$ are the same as $x^{(a)},x^{(b)}$ after removal of $x^{(a)}_\mu,x^{(b)}_\nu$, and if $y^{(c+1)}$ after removal of $y^{(c+1)}_\xi$ agrees with $x^{(c)}$, then there are $c+1$ possibilities of augmenting $x^{(c)}$ with the target position $y^{(c+1)}_\xi$ to generate a $(c+1)$-tupel. The probability that out of the reaction of reactants at $x^{(a)}_\mu,x^{(b)}_\nu$ emerges a particle in $y^{(c+1)}_\xi$ must be equi-distributed over these equivalent configurations of $c+1$ product particles to retain the required particle exchange symmetry.

Regarding the symmetrization we observe that there are $(a-1)! \, (b-1)! \, (c+1)!$ equivalent configurations in the target space over which the probability of being transferred to  has to be distributed. In analogy to \cref{eq:PropensityDecay}, we obtain

\begin{multline}\label{eq:lambda_ABC}
  \Lambda\left(y^{(a-1)},y^{(b-1)},y^{(c+1)}; x^{(a)},x^{(b)},x^{(c)}\right)
  \\[10pt]
  = \sum_{P\in\mathcal{P}_{a-1}} \sum_{Q\in\mathcal{P}_{b-1}} \sum_{R\in\mathcal{P}_{c+1}}
  \frac{
	\Lambda^{\text{bs}}\left(Py^{(a-1)},Qy^{(b-1)},Ry^{(c+1)}; x^{(a)},x^{(b)},x^{(c)}\right)}{(a-1)! \, (b-1)! \, (c+1)!}\,.
\end{multline}
The formula in \cref{eq:lambda_ABC} is obviously symmetric w.r.t.\ the target space configurations by construction. It is also symmetric w.r.t.\ the source space configurations, because of the summation over all possible pairs of reactant particles in \cref{eq:lambdaBare_ABC} and the symmetrization over the target space configurations in \cref{eq:lambda_ABC}.

Let us now derive the structure of the loss and gain expressions analogous to those in \cref{loss-gain} for this representation of the many-particle propensity.

\paragraph{The loss term $\mathcal{L}$.}

Extending the definition in \cref{eq:LDefinition} to the two-species reaction and dropping the superscript on $\mathcal{L}$ as it is clear from the context, we have
\begin{equation}\label{eq:LossR2}
(\mathcal{L}\rho_{a,b,c})\bigl(x^{(a)},x^{(b)},x^{(c)}\bigr)
= \rho_{a,b,c}\bigl(x^{(a)},x^{(b)},x^{(c)}\bigr) \
\sum_{\mu=1}^{a}\sum_{\nu=1}^{b} 
\int_{\mathbb{X}} \lambda\bigl(y;x^{(a)}_\mu,x^{(b)}_\nu\bigr) dy\,.  
\end{equation}
To obtain this result, we have used that integration over just one of the terms in the multiple sum over particle indices in \cref{eq:lambdaBare_ABC} and permutations in \cref{eq:lambda_ABC} may be summarized as follows. Dropping the prefactors of $1/(c+1)$ and $\rho_{a,b,c}(x^{(a)},x^{(b)},x^{(c)}) \big/ (a-1)! \, (b-1)! \, (c+1)!$ for the moment, we consider only the terms relevant for the integration, i.e.,
\begin{equation}
\begin{array}{l}
\dss \int\limits_{\mathbb{X}^{a-1}\times\mathbb{X}^{b-1}\times\mathbb{X}^{c+1}} \hspace{-5mm}
\Lambda_{\nu,\mu,\xi}
\left(Py^{(a-1)},Qy^{(b-1)},Ry^{(c+1)};x^{(a)},x^{(b)},x^{(c)}\right)
dy^{(a-1)}\,dy^{(b-1)}\,dy^{(c+1)}
\\[20pt]
= \dss \int\limits_{\mathbb{X}^{a-1}\times\mathbb{X}^{b-1}\times\mathbb{X}^{c+1}} \hspace{-5mm}
\Lambda_{\nu,\mu,\xi}
\left(y^{(a-1)},y^{(b-1)},y^{(c+1)}; x^{(a)},x^{(b)},x^{(c)}\right) 
dy^{(a-1)}\,dy^{(b-1)}\,dy^{(c+1)}
\\[20pt]
= \dss \int\limits_{\mathbb{X}} 
\lambda\bigl(y;x^{(a)}_\mu,x^{(b)}_\nu\bigr) 
\ dy\,  .
\end{array}
\end{equation}
Here the first equality follows by a transformation of the integration variables from the components of $\bigl(y^{(a-1)}, y^{(b-1)}, y^{(c+1)}\bigr)$ to the components of $\bigl(Py^{(a-1)}, Qy^{(b-1)}, Ry^{(c+1)}\bigr)$ and relabelling. The second equality follows because all the $\delta$-distributions in \cref{eq:lambdaBareIndividual_ABC} will generate unity once upon the integrations over the $y^{(a-1)}_i \ (i=1,\dots,a-1)$, $y^{(b-1)}_j \ (j=1,\dots,b-1)$ and $y^{(c+1)}_k \ (k = 1, \dots, \xi-1, \xi+1, c+1)$, whereas the integration over $y = y^{(c+1)}_\xi$ remains non-trivial. Thus we observe, that all these terms are identical for any of the $c+1$ terms in the sum over $\xi$ in \cref{eq:lambdaBare_ABC} and as well for any of the permutations in \cref{eq:lambda_ABC}. Carrying out the summation over the permutations yields a factor of $(a-1)!\, (b-1)!\, (c+1)!$ which cancels the denominator in \cref{eq:lambda_ABC}, while summing over $\xi$ in \cref{eq:lambdaBare_ABC} cancels the factor of $1/(c+1)$ in that equation. This establishes~\cref{eq:LossR2}.

\paragraph{The gain term  $\mathcal{G}$.}

To calculate the gain operator $\mathcal{G}$ for the target space, $\mathbb{X}^{a}\times\mathbb{X}^{b}\times\mathbb{X}^{c}$, of the reaction, we have to compute the expectation of the propensity over the source space, $\mathbb{X}^{a+1}\times\mathbb{X}^{b+1}\times\mathbb{X}^{c-1}$, in analogy with \cref{eq:GDefinition}. The associated density-weighted integration over $\bigl(x^{(a+1)},x^{(b+1)},x^{(c-1)}\bigr)$ of $\Lambda_{\nu,\mu,\xi}$ in \cref{eq:lambdaBareIndividual_ABC} yields
\begin{multline}\label{eq:ABCSingleTermIntegral}
\int\limits_{\mathbb{X}^{a}\times\mathbb{X}^{b}\times\mathbb{X}^{c}} 
\Lambda_{\nu,\mu,\xi}\left(y^{(a-1)}\!,y^{(b-1)}\!,y^{(c+1)}; x^{(a)}\!,x^{(b)}\!,x^{(c)}\right)
\rho_{a,b,c}\bigl(x^{(a)},x^{(b)},x^{(c)}\bigr)\, 
dx^{(a)}\,dx^{(b)}\,dx^{(c)}
\\[10pt]
= \int_{\mathbb{X}^2}
\lambda(y^{(c+1)}_\xi;z,z',)\
\rho_{a,b,c}\!
\left( \bigl(y^{(a-1)},z\bigr),\bigl(y^{(b-1)},z'\bigr),
y^{(c+1)}_{\setminus \{\xi\}}\right)\, 
dz \,dz'\,.
\end{multline}
Here we have already used the symmetry properties of $\rho_{a,b,c}$ to shift the remaining integration variables $z$ and $z'$ to the end of the tuples of its first two arguments.
These calculations show that the result is again independent of the summation indices $\mu,\nu$, so that the summation over these indices in \cref{eq:lambdaBare_ABC} just generates a prefactor of $a b$. Summation over $\xi$ guarantees that the configuration $\bigl(y^{(a-1)}\!,y^{(b-1)}\!,y^{(c+1)}\bigr)$ receives its appropriate share of probability transfer from all reactions that produce a particle in any of the positions collected in the tuple $y^{(c+1)}$.

Any permutation of $y^{(a-1)}$ or $y^{(b-1)}$ will not change the result either owing to the symmetry of $\rho_{a,b,c}$ in its first two arguments. Therefore, the averaging over these permutations will just cancel the prefactor of $1/(a-1)!\, (b-1)!$ in \cref{eq:lambda_ABC}. After the summation over $\xi$ in \cref{eq:lambdaBare_ABC}, the resulting expression is invariant under permutations of $y^{(c+1)}$ as well owing to the symmetry of $\rho_{a,b,c}$ in its last argument. Thus, the summation over these permutations will just generate a factor of $(c+1)!$, canceling the remaining factor in the denominator of \cref{eq:lambda_ABC}. Note, however, that the factor of $1/(c+1)$ from \cref{eq:lambdaBare_ABC} is retained in the process.

The result for the gain function reads

\begin{multline}\label{eq:GainR2}
(\mathcal{G}\rho_{a,b,c})\bigl(y^{(a-1)},y^{(b-1)},y^{(c+1)}\bigr) 
\\[10pt]
= \frac{ab}{c+1} \
\sum_{\xi=1}^{c+1} 
\int_{\mathbb{X}^2} \lambda\bigl(y^{(c+1)}_\xi;z,z'\bigr)
\rho_{a,b,c}\!
\left(\bigl[y^{(a-1)},z\bigr],\bigl[y^{(b-1)},z'\bigr],
y^{(c+1)}_{\setminus \{\xi\}}\right)\, dz\, dz'\,.
\end{multline}
After a shift from $(a,b,c)$ to $(a+1,b+1,c-1)$, we obtain an operator which agrees with the mid term in \cref{CDME:ABC}.
Preservation of total probability under the loss and gain functions in \cref{eq:LossR2} and \cref{eq:GainR2} is guaranteed as we have

\begin{align}
\lefteqn{\int\limits_{\mathbb{X}^{a}\times\mathbb{X}^{b}\times\mathbb{X}^{c}} \hspace{-5mm} \big(\mathcal{L} \rho_{a,b,c}\big)\bigl(x^{(a)},x^{(b)},x^{(c)}\bigr) \, dx^{(a)}\,dx^{(b)}\,dx^{(c)}} \nonumber \\
& \hspace{2cm} =\int\limits_{\mathbb{X}^{a-1}\times\mathbb{X}^{b-1}\times\mathbb{X}^{c+1}} \hspace{-10mm} \big(\mathcal{G}\rho_{a,b,c}\big)\bigl(y^{(a-1)},y^{(b-1)},y^{(c+1)}\bigr) \ dy^{(a-1)}\,dy^{(b-1)}\,dy^{(c+1)} \\
& \hspace{2cm} =  a\, b \int\limits_{\mathbb{X}^3} \lambda(y;z,z')
\hspace{-5mm}\int\limits_{\mathbb{X}^{a-1}\times\mathbb{X}^{b-1}\times\mathbb{X}^{c}} 
\hspace{-5mm}\rho_{a,b,c}\!
\left((\xi,z), (\eta,z'), \zeta\right)\, d\xi d\eta d\zeta \, dy dz dz'\,.
\end{align}

Collecting the loss and gain terms and adding the diffusion terms, we obtain again the CDME given by \cref{CDME:ABC}.

In total, we end up with the same equation, but the way to get there is different: In \cref{sec:CDME} we have expressed the loss and gain operators as sums of \textit{local} operators (acting on subsets of particles), while here in \cref{sec:ManyParticlePropensity} we have translated the local rate functions into many-particle propensities. In \cref{sec:FockCDME} the combinatorics will be encoded in the annihilation and creation operators, again ending up in the same CDME.

\section{CDME formulation using creation and annihilation operators} \label{sec:FockCDME}

Using creation and annihilation operators as presented in ref.~\onlinecite{del2021probabilistic}, we can formulate the CDME at once without having to worry about the combinatorial factors. Assuming a system involving only one chemical species, we introduce the creation and annihilation operators acting on an $n$-particle density $\rho_{n}$ as \cite{del2021probabilistic}
\begin{subequations}\label{eqs:creaannihops}
\begin{align}
	\Big(a^+\{w\}\rho_{n}\Big)(x^{(n+1)}) &=\frac{1}{n+1}\sum_{j=1}^{n+1} w(x_j^{(n+1)})\rho_n(x_{\setminus \{j\}}^{(n+1)}), \\
	\Big(a^-\{f\}\rho_{n}\Big)(x^{(n-1)}) &= n\int_\mathbb{X} f(y)  \rho_{n}\left(x^{(n-1)}, y\right) \, dy.
\end{align}
\end{subequations}
The creation operator $a^+\{w\}$ adds a particle of species $A$ with distribution $w$ by multiplying the single-particle density $w$ with the density $\rho_n$. The resulting density is a function of $n+1$ positions, $x^{(n+1)}$, and the sum over $j$ and the prefactor are required to render the result symmetric with respect to permutations of particle labels.
The annihilation operator $a^-\{f\}$ removes a particle at $x$ with the rate $f(x)$ by marginalization of the density with the weight function~$f$. As $\rho_n$ is symmetric, we can simply integrate against the last variable.
The resulting density is a function of $x^{(n-1)}$.
As there are $n$ possible ways to remove a particle, the factor of $n$ appears in front of the integral.
In ref.~\onlinecite{del2021probabilistic} it was shown that the creation and annihilation operators satisfy some special properties that are useful for calculations, including the commutation relations
\begin{align}
	\Big [a^-\{f\},a^+\{w\}\Big]  = \big\langle f,w \big\rangle, \ \ \ \ \
	\Big [a^-\{f\},a^-\{g\}\Big] = \Big[a^+\{w\}, a^+\{\nu\})\Big] = 0,
	\label{eq:a+a-CommutRelsFull}
\end{align}
where $	\langle u,v\rangle := \int_{\mathbb{X}}u(x)v(x) dx$ for suitable functions $u, v$ and $\big[a,b\big] := ab-ba$ for operators $a, b$.
Furthermore, the definitions of $a^+$ and $a^-$ extend naturally to the family of $n$-particle densities by
operating element-wise, e.g., $a^+\{w\} (\rho_0, \rho_1, \dots) = (a^+\{w\} \rho_0, a^+\{w\}\rho_1, \dots)$.

The following representation of the CDME will be given in terms of a basis $(u_1,u_2,\dots)$ of the space of single-particle densities. We emphasize that the obtained results are independent of the specific basis chosen, although the expansion coefficients will naturally depend on the choice of the basis.
For a concrete application, the basis functions can be adapted to the problem and reflect some physical properties, e.g., possible symmetries.\footnote{We recall that, in quantum mechanics, the common expansions in terms of spherical harmonics and associated polynomials is motivated by the isotropy of atoms.}
For keeping the presentation concise, we restrict here to square-integrable probability densities, which form a separable Hilbert space and assume that the basis is orthonormal, i.e., $\langle u_\alpha, u_\beta\rangle=\delta_{\alpha,\beta}$.
More generally, one uses the Banach space $L^1(\mathbb{X})$ of integrable functions as it was done in ref.~\onlinecite{del2021probabilistic}.
However, this adds a number of technical issues, and there are no relevant differences in the final expressions.
In both cases, the existence of a basis $(u_1, u_2, \dots)$ is granted, and in the $L^1(\mathbb{X})$ case, the representations are exact in the sense that every probability density can be expanded in such a basis.

\subsection{One species}

Let us consider again a general one-species reaction $kA \rightarrow lA$ with rate function $\lambda(y^{(l)}; x^{(k)})$.
Following \cref{eq:diffope}, the diffusion operator $\mathcal{D}_n$ decomposes into single-particle diffusions $D_\nu$ applied to the $\nu$th particle, which can can be expanded in terms of creation and annihilation operators \cite{del2021probabilistic}:
\begin{align}
	\mathcal{D}_n &= \sum_{\nu=1}^n D_\nu \\
	&= \sum_{\alpha,\beta} \left\langle u_\alpha, D u_{\beta}\right\rangle a^+_\alpha a^- _\beta,
	\label{eq:diffexpansion}
\end{align}
where we used the compressed notation $a^+_\alpha := a^+\{u_\alpha \}$ and $a^-_\beta := a^-\{u_\beta \}$.

One observes that the expansion in \cref{eq:diffexpansion} does not depend explicitly on the particle number $n$ and thus, formally, it represents the full diffusion operator acting on the whole family $\rho=(\rho_0, \rho_1, \dots)$.

We now need to expand the loss and gain operators in the same manner. First, we consider the loss and gain operators per reaction from \cref{eq:lossperreaction,eq:gainperreaction}, which are linear operators and are thus fully specified by their action on products of single-particle basis functions:
\begin{align}
	\big(L(u_{\beta_1}\otimes\dots\otimes u_{\beta_k})\big)(x^{(k)}) &:= (u_{\beta_1}\otimes\dots\otimes u_{\beta_k}) (x^{(k)})\int_{\mathbb{X}^l} \lambda(y^{(l)};x^{(k)}) dy^{(l)}, \label{eq:propOperators_k}\\
	\big(G(u_{\beta_1}\otimes\dots\otimes u_{\beta_k})\big)(y^{(l)}) &:= \int_{\mathbb{X}^k} \lambda(y^{(l)};x^{(k)})(u_{\beta_1}\otimes\dots\otimes u_{\beta_k}) (x^{(k)}) dx^{(k)} \,, \label{eq:propOperators_kl}
\end{align}
with the tensor product $v_1 \otimes \dots \otimes v_n = \bigotimes_{j=1}^{n} v_j$ defined as
$(v_1 \otimes \dots \otimes v_n)(x^{(n)}) := v_1(x^{(n)}_1) \dots v_n(x^{(n)}_n)$.
One can show that $(u_{\alpha_1} \otimes \dots \otimes u_{\alpha_n})_{\alpha_i \in \mathbb{N}}$ is a basis of the corresponding tensor space of Hilbert spaces, which is itself a Hilbert space, referred to as a Fock space.
Analogous to the diffusion operator, the total loss and gain over all possible reactions from \cref{eq:losskA-lA,eq:gainkA-lA} also have expansions in terms of creation and annihilation operators \cite{del2021probabilistic},
\begin{align}
	\mathcal{L}_n &= \frac{1}{k!}
	\sum_{\substack{\alpha_1,\dots,\alpha_k\\ \beta_1,\dots,\beta_k }} \left\langle \bigotimes_{i=1}^{k} u_{\alpha_i}, L \bigotimes_{j=1}^{k} u_{\beta_j}\right\rangle \prod_{i=1}^k a^+_{\alpha_i} \prod_{j=1}^k a^-_{\beta_j}, \label{eq:lossexpansion} \\
	\mathcal{G}_n &=\frac{1}{k!}
	\sum_{\substack{\alpha_1,\dots,\alpha_l\\ \beta_1,\dots,\beta_k }} \left\langle \bigotimes_{i=1}^{l} u_{\alpha_i}, G \bigotimes_{j=1}^{k} u_{\beta_j}\right\rangle \prod_{i=1}^l a^+_{\alpha_i} \prod_{j=1}^k a^-_{\beta_j}. \label{eq:gainexpansion}
\end{align}

\begin{figure*}
	\includegraphics[width=.8\textwidth]{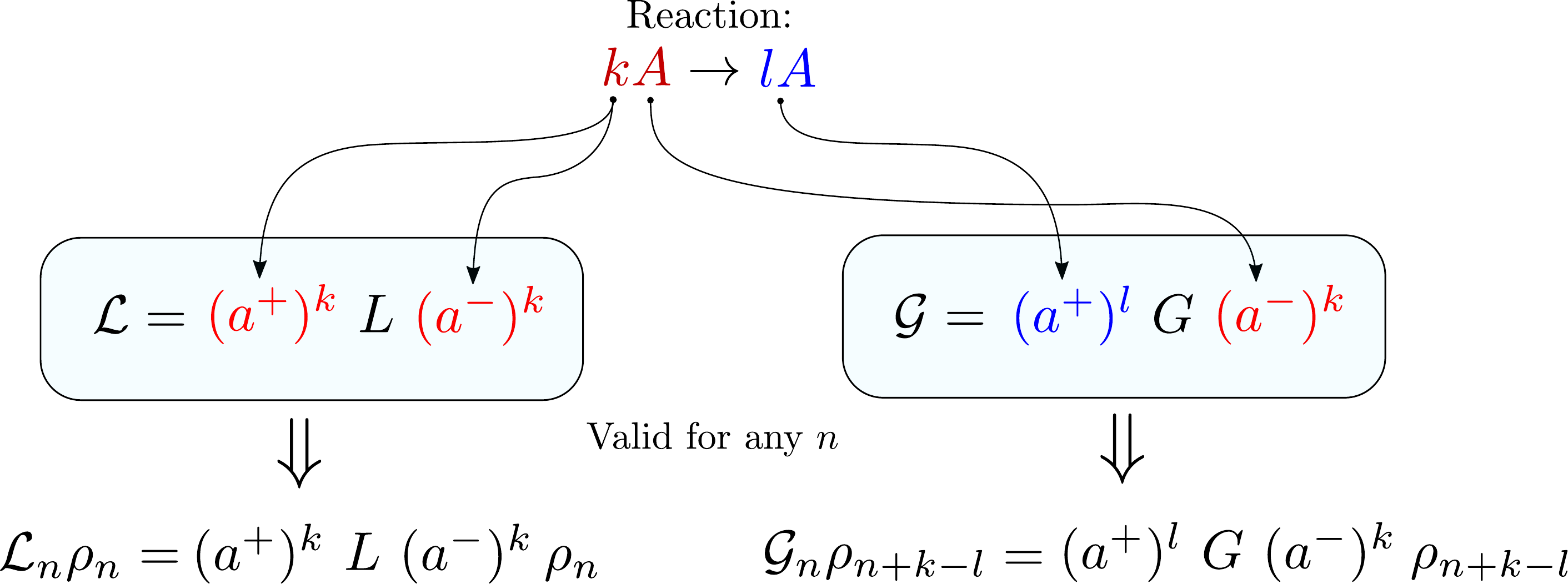}
	\caption{Diagram representing how to write the loss and gain operators for the reaction $kA\rightarrow lA$ in the compact notation using creation and annihilation operators. The operators $L$ and $G$ represent the \textit{local} loss and gain operators; the operators $\mathcal{L}$ and $\mathcal{G}$ represent the \textit{global} loss and gain operators. }
	\label{fig:secondquant}
\end{figure*}

These expansions seem to be rather involved at first sight, yet they are a key element to develop a straightforward formulation of the CDME even for complex reaction--diffusion networks. The structure of the expressions becomes more transparent by introducing the following short-hand notation.
Let $\mathrm{a}^+ = (a^+\{u_\alpha\})_{\alpha \in \mathbb{N}}$ denote the family of creation operators for the basis $(u_\alpha)$, and analogously $\mathrm{a}^- = (a^-\{u_\beta\})_{\beta \in \mathbb{N}}$.
For the coefficients of $\mathcal{D}_n$ in \cref{eq:diffexpansion}, we arrange them as
$\mathrm{D} = (\langle u_\alpha, D u_\beta\rangle)_{(\alpha,\beta)\in \mathbb{N}^2}$, which is reminiscent of a tensor of rank~2.
The expansion of $\mathcal{D}_n$ then reads
\begin{equation}
  \mathcal{D} = \contract{\mathrm{a}^+ \, \mathrm{D} \, \mathrm{a}^-} \,,
\end{equation}
where the products between the symbols in upright font face % under the bar
imply full contractions of the ``tensor'' indices $\alpha$ and $\beta$, see \cref{eq:diffexpansion};
here, we have dropped the subscript $n$ from $\mathcal{D}_n$ noting again that the right-hand side holds for any $n$.
For the loss and gain terms, we make use of multi-indices $\boldsymbol\alpha = (\alpha_1, \dots, \alpha_k)$ and write
$(\mathrm{a}^+)^k = (a^+\{u_{\alpha_1}\} \cdots a^+\{u_{\alpha_k}\})_{\boldsymbol\alpha\in\mathbb{N}^k}$
and analogously for $(\mathrm{a}^-)^l$. The coefficients of $\mathcal{L}_n$ in \cref{eq:lossexpansion} are denoted as
$\mathrm{L} = \left(
  \langle u_{\alpha_1} \otimes \cdots \otimes u_{\alpha_k}, L u_{\beta_1} \otimes \cdots \otimes u_{\beta_k} \rangle
\right)_{\boldsymbol{\alpha}\boldsymbol{\beta}}$.
With this compact notation, the expansions of the gain and loss operators in \cref{eq:lossexpansion,eq:gainexpansion} take the form (\cref{fig:secondquant})
\begin{equation}
	\mathcal{L} = \contract{(\mathrm{a}^+)^k \, \mathrm{L} \, (\mathrm{a}^-)^k} \qquad \text{and} \qquad
	\mathcal{G} = \contract{(\mathrm{a}^+)^l \, \mathrm{G} \, (\mathrm{a}^-)^k} \,,
\end{equation}
with products implying contractions over multi-indices $\boldsymbol\alpha$ and $\boldsymbol\beta$;
additionally, we agree that contractions involving several annihilation operators $(\mathrm{a}^-)^k$ introduce a factor of $k!$, corresponding to the length of the multi-index $\boldsymbol\beta$.
Then, the CDME in its compact form is
\begin{equation}
	\frac{\partial \rho}{\partial t} = \left( \contract{\mathrm{a}^+ \, \mathrm{D} \, \mathrm{a}^-}
        + \contract{(\mathrm{a}^+)^l \, \mathrm{G} \, (\mathrm{a}^-)^k} 
        - \contract{(\mathrm{a}^+)^k \, \mathrm{L} \, (\mathrm{a}^-)^k}\right) \rho.
\end{equation}
To recover the explicit form of the equation (as derived in \cref{sec:CDME,sec:ManyParticlePropensity}) one must explicitly evaluate the expressions containing the creation and annihilation operators. To circumvent these often cumbersome calculations, we provide a dictionary of the expansions for common reactions in \cref{app:dictionary}, where we can easily verify that the $n$th component of this equation matches that of \cref{CDME_n}, where the loss operator always acts on $\rho_n$, and the gain operator on $\rho_{n+k-l}$.

The compact notation has a very intuitive logic behind (\cref{fig:secondquant}): Given the reaction $kA\rightarrow lA$, the loss acts on the $k$ reactants at once, so it involves $k$ creation and $k$ annihilation operators. As the gain depends on both, reactants and products, it consists of $k$ annihilation and $l$ creation operators. The diffusion operator, as it acts on solely one particle at a time, involves only one annihilation and one creation operator. If diffusion incorporated physical pair interactions, it would act on two particles at a time, so it would involve two creation and two annihilation operators.

\subsection{Bimolecular reactions}

For reaction systems involving multiple species, it is equally easy to obtain the desired equation. We only need to use different creation and annihilation operators for each species. For examples, for the reaction
\begin{equation}
A+B\rightarrow C,
\end{equation}
with rate function $\lambda(y_C; x_A,x_B)$, where $x_A$ and $x_B$ are the locations of the reactants and $y_C$ is the location of the product, we immediately obtain
\begin{align} \label{eq:fockCDMEbim}
	\frac{\partial \rho}{\partial t} = \left(\contract{\mathrm{a}^+ \, \mathrm{D}^A \, \mathrm{a}^-}
	+ \contract{\mathrm{b}^+ \, \mathrm{D}^B \, \mathrm{b}^-}
	+ \contract{\mathrm{c}^+ \, \mathrm{D}^C \, \mathrm{c}^-}
	+ \contract{\mathrm{c}^+ \, \mathrm{G} \, \mathrm{a}^- \mathrm{b}^-}
	- \contract{\mathrm{a}^+ \mathrm{b}^+ \, \mathrm{L} \, \mathrm{a}^- \mathrm{b}^-}
		\right)\rho,
\end{align}
where $\rho$ is the family of $n$-particle densities of the form $\rho_{a,b,c}(x^{(a)},x^{(b)},x^{(c)})$ for all possible values of the particle numbers $a$, $b$, and $c$. The creation and annihilation operators for each species are denoted by the corresponding lower case letter.
The first three terms describe the diffusion of the different species; the fourth term is the total loss due to reactions; and the last term is the total gain. Note that the loss of probability will only depend on the number of reactants of the current state; thus it only contains operators for the $A$ and $B$ species. On the other hand, the gain will depend on the number of reactants in another state, as well as the products needed to bring the system to the current state. These terms have the following expansions \cite{del2021probabilistic}:
\begin{align}
	\contract{\mathrm{a}^+ \mathrm{b}^+ \, \mathrm{L} \, \mathrm{a}^- \mathrm{b}^-}
	&= \frac{1}{2}
	\sum_{\substack{\alpha_1,\alpha_2\\ \beta_1,\beta_2 }} \left\langle u_{\alpha_1} \otimes u_{\alpha_2}, L (u_{\beta_1} \otimes u_{\beta_2}) \right\rangle  a^+ \{u_{\alpha_1}\} b^+\{u_{\alpha_2} \} a^- \{u_{\beta_1}\} b^- \{u_{\beta_2}\},\\
  \contract{\mathrm{c}^+ \, \mathrm{G} \, \mathrm{a}^- \mathrm{b}^-}
	&=\frac{1}{2}
	\sum_{\substack{\alpha\\ \beta_1,\beta_2 }} \left\langle u_{\alpha}, G  (u_{\beta_1} \otimes u_{\beta_2})\right\rangle c^+\{u_{\alpha} \} a^- \{u_{\beta_1}\} b^- \{u_{\beta_2}\},
\end{align}
recalling the short-hand $a^+_{\alpha_1} = a^+\{u_{\alpha_1}\}$, etc.;
the \textit{local} loss and gain operators are given in terms of the rate function $\lambda$ as
\begin{align}
	\big( L (u_{\beta_1} \otimes u_{\beta_2})\big)(x_A,x_B) &:= (u_{\beta_1} \otimes u_{\beta_2})(x_A,x_B) \int_\mathbb{X}\lambda(y_C;x_A,x_B)dy_C,
	 \\
	 \big(G  (u_{\beta_1} \otimes u_{\beta_2})\big)(y_C) &:= \int_{\mathbb{X}^2} (u_{\beta_1} \otimes u_{\beta_2})(x_A,x_B) \lambda(y_C;x_A,x_B)dx_A dx_B,
\end{align}
in analogy to \cref{eq:propOperators_k,eq:propOperators_kl}.

Using the dictionary of \cref{app:dictionary}, it is straightforward to transform \cref{eq:fockCDMEbim} into the explicit integral notation, 
\begin{align}\begin{split}
\frac{\partial \rho_{a,b,c}}{\partial t} &=
\sum_{\mu=1}^a D^A_{\mu}\rho_{a,b,c} + \sum_{\nu=1}^b D^B_{\nu}\rho_{a,b,c} +\sum_{\xi=1}^c D^C_{\xi}\rho_{a,b,c} \\
&\qquad +\frac{(a+1)(b+1)}{c}
\sum_{\xi=1}^{c} \Bigg( 
\int_{\mathbb{X}^{2}} 
\lambda \left( x^{(c)}_{\xi};z,z' \right)
\rho_{\substack{a+1, b+1, c-1}} \left(
(x^{(a)}, z ),
(x^{(b)}, z' ),
x^{(c)}_{\setminus\{\xi \}}\right)
 dz dz'
\Bigg) \\
&\qquad -\rho_{a,b,c} \sum_{\nu_1=1}^a \sum_{\nu_2=1}^b \int_{\mathbb{X}}\lambda(y;x^{(a)}_{\nu_1}, x^{(b)}_{\nu_2}) dy
\end{split}\end{align}
which is the same as \cref{CDME:ABC}.

\subsection{Enzyme kinetics}

Before closing, we develop the CDME for a real-world example, namely the Michaelis--Menten scheme for enzyme kinetics, which consists of three reactions and involves four species:
\begin{subequations}
\begin{alignat}{3}
&R_1:\quad & E+S &\rightarrow C, \\
&R_2:\quad & C &\rightarrow E + S, \\
&R_3:\quad & C &\rightarrow E+P.
\end{alignat}
\end{subequations}
The scheme describes an enzyme $E$ that can bind a substrate molecule $S$ to form the complex $C$. This complex can either dissociate again or yield a product $P$ while releasing the original enzyme. The rate functions corresponding to these reactions are $\lambda_1(y_{C}; x_E, x_S)$, $\lambda_2(y_E, y_S; x_{C})$, and $\lambda_3(y_{E}, y_{P}; x_{C})$, respectively.
The CDME is an evolution equation for the family $\rho$ of densities of the form $\rho_{e,s,p,c}(x^{(e)},x^{(s)},x^{(p)},x^{(c)})$ for all possible values of $e,s,p$, and $c$, and it takes the form
\begin{align}
	\frac{\partial \rho}{\partial t} = \mathcal{D} \rho  + \left(\sum_{r=1}^3 \mathcal{R}^{(r)} \right)\rho
\end{align}
with the diffuson and reaction operators
\begin{subequations}
\begin{align}
  \mathcal{D} &= \contract{\mathrm{e}^+ \, \mathrm{D}^E \, \mathrm{e}^-}
    + \contract{\mathrm{s}^+ \, \mathrm{D}^S \, \mathrm{s}^-}
    + \contract{\mathrm{c}^+ \, \mathrm{D}^C \, \mathrm{c}^-}
    + \contract{\mathrm{p}^+ \, \mathrm{D}^P \, \mathrm{p}^-} \,,	 \\
	\mathcal{R}^{(1)} &=	
     \contract{\mathrm{c}^+ \, \mathrm{G}_1 \, \mathrm{e}^- \mathrm{s}^-} - \contract{\mathrm{e}^+ \mathrm{s}^+ \, \mathrm{L}_1 \, \mathrm{e}^- \mathrm{s}^-}\,, \\
	\mathcal{R}^{(2)}  &=
     \contract{\mathrm{e}^+ \mathrm{s}^+ \, \mathrm{G}_2 \, \mathrm{c}^-}  	- \contract{\mathrm{c}^+ \, \mathrm{L}_2 \, \mathrm{c}^-}\,, \\
	\mathcal{R}^{(3)}  &=	
    \contract{\mathrm{e}^+ \mathrm{p}^+ \, \mathrm{G}_3 \, \mathrm{c}^-} - \contract{\mathrm{c}^+ \, \mathrm{L}_3 \, \mathrm{c}^-}\,.
\end{align}
\end{subequations}
The expansions of the operators, as well as the corresponding loss and gain operators for each reaction, are completely analogous to the previous examples.
By virtue of \cref{app:dictionary}, we obtain the CDME in its integral notation: 
\begin{align}\begin{split}\label{enzyme_minus}
	\frac{\partial \rho_{e,s,p,c}}{\partial t} &=
	\sum_{\mu=1}^e D^E_{\mu} \rho_{e,s,p,c} + \sum_{\mu=1}^s D^S_{\mu}\rho_{e,s,p,c} + \sum_{\mu=1}^p D^P_{\mu}\rho_{e,s,p,c} + \sum_{\mu=1}^c D^C_{\mu}\rho_{e,s,p,c} \\
	&+\frac{(e+1)(s+1)}{c}
	\sum_{\xi=1}^{c} \Bigg( 
	\int_{\mathbb{X}^{2}} 
	\rho_{\substack{e+1, s+1, p, c-1}} \left(
	(x^{(e)}, z ),
	(x^{(s)}, z' ),
	x^{(p)},
	x^{(c)}_{\setminus\{\xi \}}\right)
	\lambda_1 \left( x^{(c)}_{\xi} 
	;z,z' \right) dz dz'
	\Bigg) \\
	&+\frac{(c+1)}{es}
	\sum_{\mu=1}^{e} \ \
	\sum_{\eta=1}^{s} \Bigg( \int_{\mathbb{X}} 
	\rho_{e-1, s-1, p, c+1} \left(
	x^{(e)}_{\setminus\{\mu \}}, 
	x^{(s)}_{\setminus\{\eta \}},
	x^{(p)},
	(x^{(c)}, z)\right)
	\lambda_2 \left(
	x^{(e)}_{\mu},
	x^{(s)}_{\eta} 
	; z \right) dz
	\Bigg)\\
	&+\frac{(c+1)}{ep}
	\sum_{\mu=1}^{e} \ \
	\sum_{\eta=1}^{p} \Bigg( \int_{\mathbb{X}} 
	\rho_{e-1, s, p-1, c+1} \left(
	x^{(e)}_{\setminus\{\mu \}}, 
	x^{(s)},
	x^{(p)}_{\setminus\{\eta \}},
	(x^{(c)}, z)\right)
	\lambda_3 \left(
	x^{(e)}_{\mu},
	x^{(p)}_{\eta} 
	; z \right) dz
	\Bigg) \\
		&
	-\rho_{e,s,p,c} \left( \sum_{\nu_1=1}^e \sum_{\nu_2=1}^s \int_{\mathbb{X}}\lambda_1(y;x^{(e)}_{\nu_1}, x^{(s)}_{\nu_2}) dy 
	+\sum_{\nu=1}^c \int_{\mathbb{X}^2}\lambda_2(y_1,y_2;x_{\nu}^{(c)}) dy_1 dy_2
	+\sum_{\nu=1}^c \int_{\mathbb{X}^2}\lambda_3(y_1,y_3;x_{\nu}^{(c)}) dy_1 dy_3
	\right). 
	\end{split}
\end{align}

\subsection{Non-rigorous extension to Dirac $\delta$-distributions} \label{sec:FockCDMEdeltas}

According to the definitions \eqref{eqs:creaannihops} of the annihilation and creation operators, a particle is inserted with a spatial probability density $w(x)$ and removed with a position-dependent rate function $f(x)$.
From a physics perspective, classical particles have a defined position and so it should be possible to add and delete particles at a single point $y \in \mathbb{X}$ (in this case, $w$ would correspond to a point measure).
To this end, we formally extend these operators to accept Dirac $\delta$-distributions as their arguments, ignoring here any mathematical difficulties associated with it.
For $\delta_y(x) := \delta(x-y)$, we define
\begin{subequations}	\label{eqs:creaannihopsdelta}
	\begin{align}
		(a^+\{\delta_y\}\rho_{n} )(x^{(n+1)}) &=\frac{1}{n+1}\sum_{j=1}^{n+1} \delta(x_j^{(n+1)}-y)\rho_n(x_{\setminus \{j\}}^{(n+1)}), \label{eqs:creaopsdelta}\\
		(a^-\{\delta_y\}\rho_{n})(x^{(n-1)}) &= n\int_\mathbb{X} \delta(z-y)  \rho_{n}\left(x^{(n-1)}, z\right) \, dz
		= n \,\rho_n(x^{(n-1)},y) . \label{eqs:annihopsdelta}
	\end{align}
\end{subequations}
For brevity, we will write $a^+(y) = a^+\{\delta_y\}$ and $a^-(y) = a^-\{\delta_y\}$ in the following. By direct substitution and straightforward calculations analogous to the ones in ref.~\onlinecite{del2021probabilistic}, one proves that these operators satisfy the commutation relations (see also \cref{eq:a+a-CommutRelsFull})
\begin{align}
	\left[a^-(y_1), a^+(y_2)\right]  = \delta(y_1-y_2), \qquad
	\left[a^-(y_1), a^-(y_2)\right] = \left[a^+(y_1), a^+(y_2)\right] = 0,
	\label{eq:a+a-CommutRels}
\end{align}
which agree with the corresponding expressions in quantum field theory \cite{peskin2018introduction}.

In general, for an operator $A$ acting on a single particle at position $y$, such as diffusion, or an operator $B$ acting on two particles at positions $y_1$ and $y_2$, we obtain the following representations of the corresponding Fock space operators (see also eqs. (60) and (64) in ref.~\onlinecite{del2021probabilistic}):
\begin{align}
\mathcal{A} &= \int_{\mathbb{X}\times \mathbb{X}} dx dy \, a^+(x) \, \tilde A(x;y) \, a^-(y) \,, \label{eq:A_delta}\\
\mathcal{B} &= \frac{1}{2!}\int_{\mathbb{X}^2 \times \mathbb{X}^2} dx_1 dx_2 dy_1 dy_2 \,
  a^+(x_1) a^+(x_2) \, \tilde B(x_1,x_2;y_1, y_2) \, a^-(y_1) a^-(y_2) \,.
\end{align}
As a rule of thumb, given a basis expansion such as \cref{eq:diffexpansion}, the functions $u_\alpha$ are replaced by $\delta_{x_\alpha}$, and the sums over $\alpha$ and $ \beta$ are replaced by integrals over the continuous variables $x_\alpha$ and $y_\beta$, respectively.
The integral kernels $\tilde A$ and $\tilde B$ generalize the coefficient matrices and read
$\tilde A(x;y) := \langle \delta_x, A \delta_y \rangle$,
and
$\tilde B(x_1, x_2; y_1, y_2) := \langle \delta_{(x_1, x_2)}, B \delta_{(y_1, y_2)} \rangle$,
respectively;
here, $\delta_{(x_1, \dots, x_k)}(z_1, \dots, z_k) := \delta(x_1 - z_1) \cdots \delta(x_k-z_k)$ denotes the $k$-dimensional Dirac $\delta$-distribution.

In case of a "diagonal" operator, such as the loss operator $L$, the one-particle kernel reduces to $\tilde A(x;y) = \tilde A(y) \,\delta(x-y)$ for $\tilde A(y):= \tilde A(y,y)$ and \cref{eq:A_delta} simplifies to (cf.\ eq.~(24) in ref.~\onlinecite{doi1976second}):
\begin{equation}
  \mathcal{A} = \int_\mathbb{X} dy \, a^+(y) \, \tilde A(y) \, a^-(y) \,.
  \label{eq:A_delta_diag}
\end{equation}
If $A$ is a differential operator (e.g., the diffusion operator $D$), we note that $\tilde A(x;y)$ has to be interpreted in a distributional sense:
\begin{align}
 \int dy \, \phi(y) \tilde A(x;y) &= \int dy \, \phi(y) \, \langle \delta_x, A \delta_y \rangle
   = \langle \delta_x, A \left({\textstyle\int} dy \, \phi(y) \, \delta_y \right)\rangle
   = \langle \delta_x, A \phi \rangle = (A\phi)(x)
\end{align}
for suitable test functions $\phi$.

For the \textit{global} gain and loss operators of the reaction $kA\rightarrow lA$ we apply the same rules, starting from the expansions  \eqref{eq:lossexpansion} and \eqref{eq:gainexpansion}, respectively: 
\begin{align}
	\mathcal{L} &= \frac{1}{k!}\int_{\mathbb{X}^k\times \mathbb{X}^k} dx^{(k)}dy^{(k)}
	a^+(x^{(k)}) \, \tilde L(x^{(k)}; y^{(k)}) \, a^-(y^{(k)}) \,, \label{eq:lossperdelta}
\\
  \mathcal{G} &= \frac{1}{k!}\int_{\mathbb{X}^l\times \mathbb{X}^k} dx^{(l)}dy^{(k)}
  a^+(x^{(l)}) \, \tilde G(x^{(l)}; y^{(k)}) \, a^-(y^{(k)}) \,, \label{eq:gainperdelta}
\end{align}
where $a^+(x^{(k)}) := a^+(x^{(k)}_1) \cdots a^+(x^{(k)}_k)$ yields the insertion of $k$ particles at positions $x^{(k)} = (x^{(k)}_1, \dots, x^{(k)}_k)$, and analogously $a^-(x^{(k)})$ for the removal of $k$ particles; we note that the factors in these products commute.
The coefficient functions are readily calculated from the definitions of the \textit{local} loss and gain operators, $L$ and $G$:
\begin{align}
	\tilde L(x^{(k)}; y^{(k)}) &:= \langle \delta_{x^{(k)}},L\delta_{y^{(k)}} \rangle
    \stackrel{\eqref{eq:propOperators_k}}{=} \delta(x^{(k)} - y^{(k)}) \int_{\mathbb{X}^l} dz^{(l)} \, \lambda(z^{(l)};y^{(k)}) \,, \label{tildeL}\\
  \tilde G(x^{(l)}; y^{(k)}) &:= \langle \delta_{x^{(l)}},G\delta_{y^{(k)}} \rangle
    \stackrel{\eqref{eq:propOperators_kl}}{=} \lambda(x^{(l)}; y^{(k)}) \,.
\end{align}
These results together with \cref{eq:lossperdelta,eq:gainperdelta} agree with Doi's work \cite{doi1976second}. 

For the action of products of the creation and annihilation operators, we find from \cref{eqs:creaannihopsdelta} by induction:

\begin{align}
(a^+(y^{(k)})\rho_{n-k})(x^{(n)})& = (a^+(y^{(k)}_1)a^+(y^{(k)}_{\setminus \{1\}}) \rho_n)(x^{(n)}) \notag \\
& = \frac{1}{n} \sum_{j_1=1}^{n} \delta(x^{(n)}_{j_1}-y^{(k)}_1) \left(a^+(y^{(k)}_{\setminus \{1\}}) \rho_{n-k}\right)(x^{(n)}_{\setminus \{j_1\}}) \notag \\
& = \frac{1}{n(n-1)}\sum_{j_1=1}^{n}\sum_{\substack{j_2=1 \\ j_2\neq j_1}}^n \delta(x^{(n)}_{j_1}-y^{(k)}_1) \delta(x^{(n)}_{j_2}-y^{(k)}_2)\left(a^+(y^{(k)}_{\setminus \{1,2\}}) \rho_{n-k}\right)(x^{(n)}_{\setminus \{j_1,j_2\}}) \notag \\
% & = \quad \dots \notag \\
& = \frac{(n-k)!}{n!} \sum_{j_1=1}^{n}\dots\sum_{\substack{j_k=1 \\ \mathclap{j_k\neq j_1,\dots,j_{k-1}}}}^n \delta(x^{(n)}_{j_1}-y^{(k)}_1)\dots \delta(x^{(n)}_{j_k}-y^{(k)}_k) \rho_{n-k}(x^{(n)}_{\setminus \{j_1,\dots,j_k\}}) \notag \\
& = \frac{k!(n-k)!}{n!}  \sum_{1\leq j_1<\dots<j_k\leq n} \delta(x^{(n)}_{j_1}-y^{(k)}_1)\dots \delta(x^{(n)}_{j_k}-y^{(k)}_k) \rho_{n-k}(x^{(n)}_{\setminus \{j_1,\dots,j_k\}}) \label{eqs:creaopsdelta_k}
\end{align}
and, more immediately,
\begin{align}\label{eqs:annihopsdelta_k}
(a^-(y^{(k)})\rho_{n})(x^{(n-k)}) & = \frac{n!}{(n-k)!} \rho_n(x^{(n-k)},y^{(k)}) .
\end{align}
In combination with \cref{eq:lossperdelta} and \cref{tildeL}, these results deliver the explicit form of the loss term of the CDME:
\begin{align}
	(\mathcal{L} \rho_n) (x^{(n)})&\:=\: \frac{1}{k!}\int_{\mathbb{X}^l \times \mathbb{X}^k} \lambda(z^{(l)};y^{(k)}) \bigl(a^+(y^{(k)}) a^-(y^{(k)}) \rho_n\bigr) (x^{(n)}) dz^{(l)} dy^{(k)} \notag \\
	&\stackrel{\eqref{eqs:creaopsdelta_k}}{=}  \frac{(n-k)!}{n!}  \sum_{1\leq j_1<\dots<j_k\leq n}\int_{\mathbb{X}^k} \left(\int_{\mathbb{X}^l}\lambda(z^{(l)}; y^{(k)})dz^{(l)}\right)  \notag \\
	& \qquad \qquad\times\delta(x_{j_1}^{(n)}-q_1^{(k)})\dots\delta(x_{j_k}^{(n)}-q_k^{(k)})  \bigl(a^-(y^{(k)})\rho_n\bigr) (x^{(n)}_{\setminus\{j_1, \dots, j_k\}})  dy^{(k)} \notag \\
	&\:=\: \frac{(n-k)!}{n!}  \sum_{1\leq j_1<\dots<j_k\leq n}\left(\int_{\mathbb{X}^l}\lambda(z^{(l)}; x_{j_1, \dots, j_k}^{(n)})dz^{(l)}\right)   \bigl(a^-(x_{j_1, \dots, j_k}^{(n)})\rho_n\bigr) (x^{(n)}_{\setminus\{j_1, \dots, j_k\}})   \notag \\
	&\stackrel{\eqref{eqs:annihopsdelta_k}}{=} \sum_{1\leq j_1<\dots<j_k\leq n} \left(\int_{\mathbb{X}^l}\lambda(z^{(l)}; x_{j_1, \dots, j_k}^{(n)})dz^{(l)}\right)  \rho_n (x^{(n)}) . \label{eq:lossDeltaEquiv}
\end{align}

Thereby, we have recovered \cref{eq:losskA-lA}, showing consistency between this approach and the one introduced in \cref{sec:CDME}. We can repeat this exercise for the gain operator using \cref{eq:gainperdelta},
% \begin{subequations} \label{eq:gainDeltaEquiv}
\begin{align}
	(\mathcal{G} \rho_{n+k-l}) (x^{(n)})&\:=\: \frac{1}{k!} \int_{\mathbb{X}^l \times \mathbb{X}^k}  \,  \lambda(z^{(l)}; y^{(k)}) \bigl(a^+(z^{(l)}) a^-(y^{(k)}) \rho_{n+k-l}\bigr) (x^{(n)}) \, dz^{(l)} dy^{(k)}  \notag \\
	&\stackrel{\eqref{eqs:creaopsdelta_k}}{=}  \frac{l!(n-l)!}{n!k!}  \sum_{1\leq j_1<\dots<j_l\leq n}\int_{\mathbb{X}^l \times \mathbb{X}^k}\lambda(z^{(l)}; y^{(k)}) \notag \\
& \qquad \qquad\times\delta(x_{j_1}^{(n)}-z^{(l)}_1)\dots\delta(x_{j_l}^{(n)}-z^{(l)_l})  \bigl(a^-(y^{(k)})\rho_{n+k-l}\bigr) (x^{(n)}_{\setminus\{j_1\dots j_l\}}) \, dz^{(l)} dy^{(k)} \notag \\
	&\:=\: \frac{l!(n-l)!}{n!k!}  \sum_{1\leq j_1<\dots<j_l\leq n}\int_{\mathbb{X}^k} \lambda(x^{(n)}_{j_1,\dots,j_l}; y^{(k)})   \bigl(a^-(y^{(k)})\rho_{n+k-l}\bigr) (x^{(n)}_{\setminus\{j_1\dots j_l\}})  \, dy^{(k)} \notag \\
		&\stackrel{\eqref{eqs:annihopsdelta_k}}{=} \frac{l!(n+k-l)!}{n!k!}  \sum_{1\leq j_1<\dots<j_l\leq n}\int_{\mathbb{X}^k} \lambda(x^{(n)}_{j_1,\dots,j_l}; y^{(k)})  \rho_{n+k-l} (x^{(n)}_{\setminus\{j_1\dots j_l\}},y^{(k)})  \, dy^{(k)} \notag \\
		&\:=\: \binom{n}{l}^{-1} \binom{n+k-l}{k} \sum_{1\leq j_1<\dots<j_l\leq n}\int_{\mathbb{X}^k} \lambda(x^{(n)}_{j_1,\dots,j_l}; y^{(k)})  \rho_{n+k-l} (x^{(n)}_{\setminus\{j_1\dots j_l\}}, y^{(k)}) \,  dy^{(k)} \,, \label{eq:gainDeltaEquiv}
\end{align}

once again, recovering \cref{eq:gainkA-lA} from \cref{sec:CDME}. The relations in the dictionary from \cref{app:dictionary} are proved in a similar fashion, but using the expansions of \cref{sec:FockCDME} as shown in ref.~\onlinecite{del2021probabilistic}.

We can further obtain a relation between the rate functions and the many-particle propensities by comparing the resulting loss from \cref{eq:lossDeltaEquiv} with the many particle propensity in \cref{eq:LDefinition}. This relation holds regardless of the density,
\begin{align}
	\int_{\mathbb{X}^{n-k+l}} \Lambda_n(y^{(n-k+l)};x^{(n)}) dy^{(n-k+l)} = \sum_{1\leq j_1<\dots<j_k\leq n}  \int_{\mathbb{X}^{l}}\lambda(y^{(l)}; x_{j_1, \dots, j_k}^{(n)})dy^{(l)}.
\end{align}
This establishes a connection with \cref{sec:ManyParticlePropensity}. We can prove this identity independently by deriving the expressions of the many-particle propensities for the reaction $kA\rightarrow lA$.

\section{Discussion}
We presented three approaches to formulate the CDME, the governing equation of stochastic particle-based reaction--diffusion dynamics. In general, the CDME consists of a diffusion operator, which  describes the spatial transport of particles, and several reaction operators each corresponding to a chemical reaction in the system. Every reaction operator can further be separated into a loss and a gain operator for the probabilistic outflow and inflow, respectively. 

In the first approach, these \textit{global} loss and gain operators have been expressed as combinations of \textit{local} loss and gain operators referring to reactions of subsets of reactants and products within the system. The central combinatorial factors, which come into play due to the particle exchange symmetry for molecules of the same species, have been justified by carefully applying combinatorical arguments for the random selection of subsets of particles out of a larger set. Although this approach is intuitive and relatively straightforward, it requires computing the combinatorial factors of the reaction operators by hand, and it is error-prone when writing the equations for complicated systems. 

The second approach (\cref{sec:ManyParticlePropensity}) works directly at the many-particle level by focusing on many-particle propensities, leaving the counting/combinatorial details as a secondary task, albeit still a cumbersome one. The global many-particle propensities are derived as explicit expressions (in terms of sums and products) of the local rate functions using permutations and Dirac $\delta$-distributions, which provide a method to select the required particles. One of its main advantages is that, as it works directly with many-particle propensities, it is capable of incorporating crowding effects in a more straightforward manner than the other approaches. 

In the third approach (\cref{sec:FockCDME}), the operators arise in the form of expansions containing single-particle creation and annihilation operators, which encode the combinatorics of particle selections. This allows us to focus on formulating only the operators per reaction, yielding a fast method to write down the CDME in a compact way for any reaction system, which can be a big advantage from a practical point of view. The resulting equation can be employed to perform analytical calculations, for instance one can directly apply Galerkin discretizations \cite{del2021probabilistic,froemberg2021generalized}, opening the door for ready-to-use software libraries for numerical implementation, as well as to apply methodologies from quantum field theory \cite{doi1976stochastic,grassberger1980fock}. In addition, the actions of the operators $a^+$ and $a^-$ have immediate interpretations within the stochastic Malliavin calculus \cite{lanconelli2022using}, which may open a new perspective on the stochastic description of reaction--diffusion systems.
However, the compact version of the CDME can appear obscure for practitioners used to more classical formulations in terms of integrals. To mitigate this issue, we added a dictionary (see \cref{app:dictionary}) to translate the short-hand notation for expansions in terms of creation and annihilation operators to concrete algebraic expressions which explicitly include the combinatorial factors, sums and integrals. This could be further automatized using a symbolic algebra software. We finally explore a special case using $\delta$-distributions (\cref{sec:FockCDMEdeltas}), which simplify the original expansions into simple integrals. Although the ease to derive discretizations --as well as some mathematical rigor-- is lost, some practitioners might find this approach more suitable.
 
From a mathematical perspective, the CDME is formulated in terms of density functions. Another question of interest for future research is how to formulate a corresponding equation in terms of probability measures as in ref.~\onlinecite{belavkin2003general}. This is of relevance since such a formulation might be more familiar to some mathematicians working on tangential fields, where one requires analogous models to reaction--diffusion, such as social dynamics \cite{conrad2018mathematical,djurdjevac2018human,veloz2014reaction}.

One of the main future prospective applications of the CDME is to unify most of the well-known reaction--diffusion models at different
scales, establishing the relationships between them and yielding a theoretical and computational framework for multiscale modeling of biochemical reaction systems.  For instance, we believe that the well-known models of diffusion-influenced reactions \cite{agmon1990theory,collins1949diffusion,doi1976stochastic,hanggi1990reaction,smoluchowski1918versuch,szabo1980first, szabo1989theory}, as well as recent developments \cite{gopich2018theory, gopich2019diffusion, dibak2019diffusion}, can be recovered as special cases of the CDME.
Moreover, in refs.~\onlinecite{delrazo2016discrete,del2018grand, kostre2021coupling}, macroscopic reaction--diffusion models emerge as the large copy number limit of the corresponding particle-based models, the latter being special cases or discretizations of the CDME. This further yields a precise relation between the macroscopic parameters and those at the particle level, allowing for consistent multiscale simulations \cite{flegg2012two,kostre2021coupling}.
Another example is given by a recent simulation scheme to couple Markov models of molecular kinetics with particle-based reaction--diffusion simulations \cite{dibak2018msm, del2021multiscale}, where the root model used to derive the schemes is once again a special case of the CDME. Similarly, in ref.~\onlinecite{chen2014brownian}, the authors use a hierarchy of Fokker--Planck equations to model the variable number of ions in an ion channel; a model which we also believe is a special case of the CDME. All in all, the CDME has the potential to unify a diverse range of reaction--diffusion models at different scales, yielding mathematical relationships that serve as the key ingredient to derive novel hybrid multiscale simulations for biochemical dynamics that capture the cascades of interactions across scales.

\begin{acknowledgments}
We acknowledge the support of Deutsche Forschungsgemeinschaft (DFG) through the Collaborative Research Center SFB~1114 “Scaling Cascades in Complex Systems”, project no.~235221301, sub-projects C01 and C03, and under Germany's Excellence Strategy -- MATH+ : The Berlin Mathematics Research Center (EXC-2046/1) -- project no.\ 390685689 (subproject AA1-1). MJR acknowledges support from DFG grant no. RA 3601/1-1 and from the Dutch Institute for Emergent Phenomena (DIEP) cluster at the University of Amsterdam.
\end{acknowledgments}

\appendix

\section{Expansion dictionary}
\label{app:dictionary}
Although using the notation presented in \cref{sec:FockCDME} results in writing the CDME at once, it is not evident to find the connection to the more classical form of the equation. In this appendix, we present a dictionary for the most used cases, where we match the expansions in terms of creation and annihilation operators in compact notation with their corresponding expressions in explicit integral form. These expressions, although non-trivial, are straightforward to prove along the lines given in ref.~\onlinecite{del2021probabilistic}.
At first, we present the expansions for the diffusion, then for loss operators, where the form is simpler as compared to the gain terms because it only depends on the reactants. Finally, we proceed with the gain operators. For the purpose of generality, we use the notation $\rho_{\dots}$ with the dots in the subindex indicating the unknown species involved in the reaction, e.g., we write  $\rho_{a,b,\dots}(x^{(a)},x^{(b)},\dots)$, where the dots represent numbers and positions of other species, respectively.

\subsection{Diffusion operators}
In the absence of physical interactions, the diffusion operators only act on one particle at a time, so they are the most simple ones:
\begin{align}
	\contract{\mathrm{a}^+ \, \mathrm{D} \, \mathrm{a}^-} \, \rho_{n,\dots} = \sum_{\nu=1}^n D_\nu \rho_{n,\dots}.
\end{align}

\subsection{Loss operators}
For the loss operators only the reactants are relevant, while the products just determine the variables of integration. Thus, we denote the positions of the $l$ products by $y^{(l)}$, regardless of their species. 

\paragraph*{\textbf{(i) Reactions of the form $\emptyset \rightarrow$ ($l$ products).}}
The reaction rate function is given by $\lambda(y^{(l)};)$. Here, we leave the semicolon inside the rate function in order to emphasize that there are no reactants. The expression is simply given by
\begin{align}
	\contract{\mathrm{1} \, \mathrm{L} \, \mathrm{1}} \, \rho_{\dots} = \rho_{\dots}\int_{\mathbb{X}^l}\lambda(y^{(l)};) dy^{(l)}.
\end{align}

\paragraph*{\textbf{(ii) Reactions of the form $A \rightarrow$ ($l$ products).}}
The reaction rate function is given by $\lambda(y^{(l)};x)$. Let the number of $A$-particles be $n$, and denote by $x^{(n)}_1,\dots,x^{(n)}_n$ the positions of the $n$ possible reactants. Then
\begin{align}
	\left(\contract{\mathrm{a}^+ \, \mathrm{L} \, \mathrm{a}^-} \, \rho_{n,\dots}\right)(x^{(n)},\dots) = \rho_{n,\dots}(x^{(n)},\dots) \sum_{\nu=1}^n \int_{\mathbb{X}^l}\lambda(y^{(l)};x_{\nu}^{(n)}) dy^{(l)}.
\end{align}

\paragraph*{\textbf{(iii) Reactions of the form $A + A \rightarrow$ ($l$ products).}}
The reaction rate function is given by $\lambda(y^{(l)};x_1,x_2)$, where $x_1$ and $x_2$ are the positions of the reactants. Then
\begin{align}
	\left(\contract{(\mathrm{a}^+)^2 \, \mathrm{L} \, (\mathrm{a}^-)^2} \, \rho_{n,\dots}\right)(x^{(n)},\dots) = \rho_{n,\dots}(x^{(n)},\dots) \sum_{1\leq \nu_1 < \nu_2\leq n} \int_{\mathbb{X}^l}\lambda(y^{(l)};x_{\nu_1}^{(n)}, x_{\nu_2}^{(n)}) dy^{(l)},
\end{align}
where again $x^{(n)}_1,\dots,x^{(n)}_n$ denote the positions of the $n$ possible reactants.

\paragraph*{\textbf{(iv) Reactions of the form $A + B \rightarrow$ ($l$ products).}}
The reaction rate function is given by $\lambda(y^{(l)};x,z)$, where $x$ is the position of the $A$ reactant and $z$ is the position of the $B$ reactant. Let $a$ and $b$ be the numbers of $A$ and $B$ particles, as well as $x^{(a)}_1,\dots,x^{(a)}_a$ and $x^{(b)}_1,\dots,x^{(b)}_b$ their positions, respectively. Then
\begin{multline}
	\left(\contract{\mathrm{a}^+ \mathrm{b}^+ \, \mathrm{L} \, \mathrm{a}^- \mathrm{b}^-} \,\rho_{a,b,\dots}\right)(x^{(a)},x^{(b)},\dots) \\
	=  \rho_{a,b,\dots}(x^{(a)},x^{(b)},\dots) \sum_{\nu_1=1}^a \sum_{\nu_2=1}^b \int_{\mathbb{X}^l}\lambda(y^{(l)};x^{(a)}_{\nu_1}, x^{(b)}_{\nu_2}) dy^{(l)}.
\end{multline}

\paragraph*{\textbf{(v) Reactions of the form $k_1 A + k_2 B \rightarrow$ ($l$ products).}}
As a generalization of all the previous examples, we can write the loss for an arbitrary reaction involving two species in their reactants. The reaction rate function is given by $\lambda(y^{(l)};x^{(k_1)},z^{(k_2)})$, where $x^{(k_1)}$ are the positions of the $A$-reactants and $z^{(k_2)}$ the positions of the $B$-reactants; $a$ and $b$ are the numbers of $A$ and $B$ particles, respectively. Then
\begin{multline}
	\left(\contract{ (\mathrm{a}^+)^{k_1} (\mathrm{b}^+)^{k_2} \, \mathrm{L} \, (\mathrm{a}^-)^{k_1} (\mathrm{b}^-)^{k_2} } \, \rho_{a,b,\dots}\right)(x^{(a)},x^{(b)},\dots) \\= \rho_{a,b,\dots}(x^{(a)},x^{(b)},\dots) \sum_{\substack{1\leq \nu_1 < \dots< \nu_{k_1}\leq a \\ 1\leq \mu_1 < \dots < \mu_{k_2}\leq b}} \int_{\mathbb{X}^l}\lambda(y^{(l)};x_{\nu_1,\dots,\nu_{k_1}}^{(a)}, x_{\mu_1,\dots,\mu_{k_2}}^{(b)}) dy^{(l)}
\end{multline}
where $x_{\nu_1,\dots,\nu_{k_1}}^{(n)}:=(x^{(n)}_{\nu_1},\dots,x^{(n)}_{\nu_{k_1}})$.

\subsection{Gain operators}
For the gain operators, both the reactants and the products are relevant, so we need to take both into account.
Once again, as the number of species will in general not be known, we indicate particle numbers and position arguments referring to non-participating species by an ellipsis, $\dots$.

\paragraph*{\textbf{(i) Reactions of the form $k_1A + k_2B \rightarrow l_1 A + l_2 B$.}}
The reaction rate function is given by $\lambda(y_A^{(l_1)},y_B^{(l_2)};x_A^{(k_1)},x_B^{(k_2)})$; $a$ and $b$ are the numbers of $A$ and $B$ particles, respectively. The expression for the gain is then
\begin{multline}
	\left(\contract{(\mathrm{a}^+)^{l_1} (\mathrm{b}^+)^{l_2} \, \mathrm{G} \, (\mathrm{a}^-)^{k_1} (\mathrm{b}^-)^{k_2}} \, \rho_{a+k_1-l_1, b+k_2-l_2,\dots}\right)(x^{(a)},x^{(b)},\dots) = \\
	C_{ab}
    \sum_{\substack{ 1\leq \mu_1< \dots< \mu_{l_1}\leq a \\ 1\leq \eta_1< \dots< \eta_{l_2}\leq b}}
    \int_{\mathbb{X}^{k_1}\times \mathbb{X}^{k_2}}
    \rho_{a+k_1-l_1, b+k_2-l_2,\dots}\mleft(
      (x^{(a)}_{\setminus\{\mu_1,\dots,\mu_{l_1} \}}, z^{(k_1)} ),
      (x^{(b)}_{\setminus\{\eta_1,\dots,\eta_{l_2} \}}, \hat z^{(k_2)} ), \dots
    \mright) \\
	\times \lambda \mleft(
    x^{(a)}_{\mu_1,\dots,\mu_{l_1}}, x^{(b)}_{\eta_1,\dots,\eta_{l_2}} ; z^{(k_1)},\hat z^{(k_2)} \mright)
    dz^{(k_1)} d\hat z^{(k_2)}
\end{multline}
with the combinatorial factor
\begin{equation}
  C_{ab}=\binom{a}{l_1}^{-1}\binom{b}{l_2}^{-1} \binom{a+k_1-l_1}{k_1}\binom{b+k_2-l_2}{k_2}
 \,.
\end{equation}

\paragraph*{\textbf{(ii) Reactions of the form $k_1A + k_2B + C\rightarrow l_1 A + l_2 B$.}}
The reaction rate function is given by $\lambda(y_A^{(l_1)},y_B^{(l_2)};x_A^{(k_1)},x_B^{(k_2)},x_C)$; $a$, $b$ and $c$ are the numbers of $A$, $B$ and $C$ particles, respectively. The expression for the gain is then
\begin{multline}
 	\left\contract{ (\mathrm{a}^+)^{l_1} (\mathrm{b}^+)^{l_2} \, \mathrm{G} \, (\mathrm{a}^-)^{k_1} (\mathrm{b}^-)^{k_2} \mathrm{c}^- } \, \rho_{a+k_1-l_1, b+k_2-l_2, c+1,\dots}\right)(x^{(a)},x^{(b)},x^{(c)},\dots) \\
 	=
    C_{abc}
    \sum_{\substack{ 1\leq \mu_1< \dots< \mu_{l_1}\leq a \\ 1\leq \eta_1< \dots< \eta_{l_2}\leq b}}
    \int_{\mathbb{X}^{k_1}\times \mathbb{X}^{k_2}\times \mathbb{X}} %\\
    \rho_{a+k_1-l_1, b+k_2-l_2, c+1,\dots} \mleft(
      (x^{(a)}_{\setminus\{\mu_1,\dots,\mu_{l_1} \}}, z^{(k_1)}),
      (x^{(b)}_{\setminus\{\eta_1,\dots,\eta_{l_2} \}}, \hat z^{(k_2)}),
      (x^{(c)}, z'), \dots
    \mright) \\
	\times \lambda \mleft(
    x^{(a)}_{\mu_1,\dots,\mu_{l_1}}, x^{(b)}_{\eta_1,\dots,\eta_{l_2}}; z^{(k_1)}, \hat z^{(k_2)}, z'
  \mright)
	dz^{(k_1)} d\hat z^{(k_2)} dz'
\end{multline}
with
\begin{equation}
  C_{abc} = \binom{a}{l_1}^{-1}\binom{b}{l_2}^{-1} \binom{a+k_1-l_1}{k_1}\binom{b+k_2-l_2}{k_2}\binom{c+1}{1} \,.
\end{equation}

\paragraph*{\textbf{(iii) Reactions of the form $k_1A + k_2B\rightarrow l_1 A + l_2 B + C$.}}
The reaction rate function is given by 
$\lambda(y_A^{(l_1)},y_B^{(l_2)}, y_C;x_A^{(k_1)},x_B^{(k_2)})$; $a$, $b$ and $c$ are the numbers of $A$, $B$ and $C$ particles, respectively. The expression for the gain is then
\begin{multline}
	\left(\contract{(\mathrm{a}^+)^{l_1} (\mathrm{b}^+)^{l_2} \mathrm{c}^+ \, \mathrm{G} \, (\mathrm{a}^-)^{k_1} (\mathrm{b}^-)^{k_2}}
    \rho_{a+k_1-l_1, b+k_2-l_2, c-1,\dots}\right)(x^{(a)},x^{(b)},x^{(c)},\dots) \\
     =
  \tilde{C}_{abc}
   \sum_{\substack{ 1\leq \mu_1< \dots< \mu_{l_1}\leq a \\ 1\leq \eta_1< \dots< \eta_{l_2}\leq b}} \ 
	\sum_{\xi=1}^{c}
	\int_{\mathbb{X}^{k_1}\times \mathbb{X}^{k_2}} 
	\rho_{a+k_1-l_1, b+k_2-l_2, c-1,\dots} \mleft(
    (x^{(a)}_{\setminus\{\mu_1,\dots,\mu_{l_1} \}}, z^{(k_1)}),
    (x^{(b)}_{\setminus\{\eta_1,\dots,\eta_{l_2} \}}, \hat z^{(k_2)}),
    x^{(c)}_{\setminus\{\xi \}}, \dots
	\mright) \\
	\times \lambda \mleft(
    x^{(a)}_{\mu_1,\dots,\mu_{l_1}}, x^{(b)}_{\eta_1,\dots,\eta_{l_2}}, x^{(c)}_{\xi}; z^{(k_1)}, \hat z^{(k_2)}
	\mright) dz^{(k_1)} d\hat z^{(k_2)}
\end{multline}
with
\begin{equation}
  \tilde{C}_{abc}=\frac{1}{c}\binom{a}{l_1}^{-1}\binom{b}{l_2}^{-1}\binom{a+k_1-l_1}{k_1}\binom{b+k_2-l_2}{k_2} \,.
\end{equation}

\paragraph*{\textbf{(iv) Reactions of the form $ C\rightarrow A +  B$.}}
This is a special case of example (ii), putting $k_1 = k_2 = 0$ and $l_1 = l_2 = 1$. The reaction rate function is given by $\lambda(y_A,y_B;x_C)$; $a$, $b$ and $c$ are the numbers of $A$, $B$ and $C$ particles, respectively. The expression for the gain is then
\begin{multline}
	\left(\contract{\mathrm{a}^+ \mathrm{b}^+ \, \mathrm{G} \, \mathrm{c}^-}
	\rho_{a-1, b-1, c+1,\dots}\right)(x^{(a)},x^{(b)},x^{(c)},\dots) = \\
	\frac{c+1}{ab}
	\sum_{\mu=1}^{a}
  \sum_{\eta=1}^{b} \int_{\mathbb{X}}
	\rho_{a-1, b-1, c+1,\dots} \mleft(
    x^{(a)}_{\setminus\{\mu \}},
    x^{(b)}_{\setminus\{\eta \}},
    (x^{(c)}, z), \dots
	\mright)
	\lambda \mleft(
    x^{(a)}_{\mu},x^{(b)}_{\eta}; z
  \mright) dz \,.
\end{multline}

\paragraph*{\textbf{(v) Reactions of the form $A + B\rightarrow  C$.}}
This is a special case of example (iii) with $k_1 = k_2 = 1$ and $l_1 = l_2 = 0$. The reaction rate function is given by $\lambda( y_C;x_A,x_B)$; $a$, $b$ and $c$ are the numbers of $A$, $B$ and $C$ particles, respectively. The expression for the gain is then
\begin{multline}
	\left(\contract{\mathrm{c}^+ \, \mathrm{G} \, \mathrm{a}^- \mathrm{b}^-}
	\rho_{a+1, b+1, c-1,\dots}\right)(x^{(a)},x^{(b)},x^{(c)},\dots) = \\
	\frac{(a+1)(b+1)}{c}
 	\sum_{\xi=1}^{c} \int_{\mathbb{X}\times \mathbb{X}}
	\rho_{a+1, b+1, c-1,\dots} \mleft(
    (x^{(a)}, z),
    (x^{(b)}, \hat z),
    x^{(c)}_{\setminus\{\xi \}}
	\dots\mright)
	\lambda \mleft(	x^{(c)}_{\xi}; z,\hat z \mright) dz d\hat z  \,.
\end{multline}

\bibliography{references}

\end{document}